\documentclass[12pt]{article}
\usepackage{amsmath,amssymb,epsfig}
\usepackage{cancel}
\usepackage{color}
\usepackage{bm}
\usepackage{braket}
\usepackage{slashed}
\usepackage{graphicx}
\usepackage{comment}
\usepackage{ulem}


\newcommand{\im}{{\rm Im\,}}
\newcommand{\re}{{\rm Re\,}}
\newcommand{\TWC }[4]{\vartheta
\begin{bmatrix}
#1\\
#2
\end{bmatrix}
\left(#3,#4\right)}

\begin{document}

\title{\vbox{
\baselineskip 14pt
\hfill \hbox{\normalsize }
} \vskip 1.7cm
\bf 
Modular Flavor Symmetry \\ on  
Magnetized Torus
\vskip 0.5cm
}
\author{
Hiroshi~Ohki$^{1}$,
Shohei~Uemura$^{2}$, \ and \
Risa~Watanabe$^{1}$
\\*[20pt]
{\it \normalsize 
${}^{1}$Department of Physics, Nara Women's University, 
Nara 630-8506, Japan}
\\
{\it \normalsize 
${}^{2}$CORE of STEM, Nara Women's University, 
Nara 630-8506, Japan}
\\*[50pt]}

\date{
\centerline{\small \bf Abstract}
\begin{minipage}{0.9\linewidth}
\medskip 
\medskip
\small
We study the modular invariance in magnetized torus models.
Modular invariant flavor model is a recently proposed hypothesis for solving the flavor puzzle, 
where the flavor symmetry originates from modular invariance.
In this framework coupling constants such as Yukawa couplings 
are also transformed under the flavor symmetry.
We show that the low-energy effective theory of magnetized torus models is 
invariant under a specific subgroup of the modular group.
Since Yukawa couplings as well as chiral zero-modes transform under the modular group, 
the above modular subgroup (referred to as modular flavor symmetry) 
provides a new type of modular invariant flavor models 
with $D_4 \times \mathbb{Z}_2$, $(\mathbb{Z}_4 \times \mathbb{Z}_2) 
\rtimes \mathbb{Z}_2$, and $(\mathbb{Z}_8 \times \mathbb{Z}_2) \rtimes \mathbb{Z}_2$.
We also find that conventional discrete flavor symmetries 
which arise in magnetized torus model 
are noncommutative with the modular flavor symmetry.
Combining both two symmetries we obtain a larger flavor symmetry, 
which is the semidirect product of the conventional flavor symmetry and the modular flavor symmetry 
for nonvanishing Wilson line.
For the vanishing Wilson line, we have additional $\mathbb{Z}_2$ symmetry, i.e., parity,
which is the unique common element between the conventional flavor symmetry 
and the modular flavor symmetry.
\end{minipage}
}


\newpage
\begin{titlepage}
\maketitle
\thispagestyle{empty}
\clearpage
\thispagestyle{empty}
\end{titlepage}

\renewcommand{\thefootnote}{\arabic{footnote}}

\section{Introduction}

The origin of the flavor structure of the quarks and leptons is a long-standing problem.
Discrete flavor symmetry is an attractive candidate answer for the flavor puzzle especially for the neutrino sector.
For instance, small $\theta_{13}$ and large $\theta_{23}$ might imply the tribimaximal mixing \cite{Harrison:2002er},
and such a characteristic pattern can be originated from discrete symmetry 
\cite{Harrison:2003aw, deMedeirosVarzielas:2006fc, Altarelli:2010gt}.
For review, see \cite{Ishimori:2010au, King:2013eh} and references therein.\,\footnote{
Recent developments of neutrino oscillation experiments unveil the precise structure of the mixing angles 
including the CP-phase  \cite{Abe:2018wpn}. }

The modular invariant flavor model is a new hypothesis proposed for solving the flavor puzzle \cite{Feruglio:2017spp, deAdelhartToorop:2011re},
which assumes that the action is invariant under the modular group $\Gamma = PSL(2,\, Z) = SL(2,Z)/\mathbb{Z}_2$.
The most distinct feature of this framework is that not only the fields, such as the leptons and the Higgs field, but also 
the coupling parameters are transformed under the modular group. 
More precisely, they form representation of quotient groups of the modular group:
$\Gamma_N=\Gamma/\Gamma(N)$.
$\Gamma_N$ is called finite modular group.
The experimental values corresponding to the lepton sectors,
the masses of charged leptons, neutrino mass-square differences, three mixing angles, 
and the CP-phase can be reproduced 
in models with modular symmetries of $\Gamma_2 \cong S_3$ \cite{Kobayashi:2019rzp, Okada:2019xqk, Kobayashi:2018vbk},
$\Gamma_3 \cong A_4$ \cite{Kobayashi:2018vbk, Kobayashi:2018scp, Criado:2018thu, Novichkov:2018yse, King:2020qaj}, $\Gamma_4 \cong S_4$ \cite{Penedo:2018nmg, Novichkov:2018ovf}, and $\Gamma_5 \cong A_5$ \cite{Novichkov:2018nkm}.
Modular symmetry is also applied to other physics beyond the standard model such as leptogenesis and inflation \cite{Asaka:2019vev, Wang:2019ovr, Kobayashi:2016mzg,Kobayashi:2018wkl},
and relationships between generalized CP symmetry \cite{Feruglio:2012cw,Holthausen:2012dk}
 and the modular symmetry are also pointed out \cite{Nilles:2018wex, Baur:2019iai, Baur:2019kwi, Novichkov:2019sqv}.

Modular symmetry is motivated by string compactifications.
So far, the modular symmetries were investigated in the heterotic string on orbifolds~\cite{Hamidi:1986vh, Dixon:1986qv, Lauer:1989ax, Lerche:1989cs, Ferrara:1989bc},  
and in the D-brane modes~\cite{Cremades:2003qj, Blumenhagen:2005mu, Abel:2006yk, Blumenhagen:2006ci}.
The situation is different in the case of 
type II superstring with magnetic flux \cite{Blumenhagen:2000wh}.
The K\"ahler potential of type IIB superstring implies that
the chiral superfield has modular weights \cite{Grana:2003ek}.
The zero-mode's profiles of bulk fields have also been investigated 
using the four-dimensional effective action compactified on torus with magnetic flux~\cite{Cremades:2004wa}.
Yukawa couplings are then obtained through the overlap integrals 
of the zero-mode wave functions.
These results have been used to investigate the property of the modular transformation for each component 
\cite{Kobayashi:2016ovu, Kobayashi:2018bff, Kobayashi:2017dyu, Kobayashi:2018rad, Kikuchi:2020frp}, 
and it is found that the Yukawa couplings as well as the chiral zero-modes form a representation of the modular group.
However it still remains unclear whether the full effective action including the Yukawa term is modular invariant. 
The purpose of this paper is to study modular invariance of the effective action of the magnetized torus model 
in a systematic way based on the fundamental generators $S$ and $T$ of the modular transformation.
We show that although the effective action is not invariant under the modular group, 
it is invariant under its specific subgroup.
The generators of the Yukawa invariant modular subgroup form a new type of flavor symmetry referred to as
modular flavor symmetry, such as $\mathbb{Z}_2$, $D_4 \times \mathbb{Z}_2$, $(\mathbb{Z}_4 \times \mathbb{Z}_2) \rtimes \mathbb{Z}_2$, 
and $(\mathbb{Z}_8 \times \mathbb{Z}_2) \rtimes \mathbb{Z}_2$ depending on the value of magnetic fluxes.
The modular flavor symmetry is noncommutative with conventional discrete flavor symmetries,
e.g., $\Delta(27)$, which appear if the greatest common divisor of generation numbers of matter fields $g$ 
is greater than 1 \cite{Abe:2009vi}.
Combing these two groups, we obtain a larger flavor symmetry.
This idea has already been discussed in \cite{Baur:2019kwi, Nilles:2020nnc, Nilles:2020kgo, Nilles:2020tdp},
in which a possible extension of the conventional flavor groups by finite modular groups has been studied in the heterotic orbifold.
In this paper,
we develop a similar idea for magnetized torus.
We find that it is insufficient for determining the group structure correctly by a single field because its representation is not faithful in the combined two groups.
To avoid this ambiguity, we consider a simultaneous transformation of all the components in the model.
We find that the conventional discrete flavor group is a normal subgroup of the whole group.
In other words the modular group is interpreted as a subgroup of the automorphism of the conventional flavor group.
This is consistent with the result of  \cite{Nilles:2020nnc}.
We also find that the whole symmetry group is isomorphic to the semidirect product of modular and the conventional flavor group
if the Yukawa couplings have a faithful representation.%
\footnote{This is not always true for the combined symmetry.
For the heterotic orbifold, the group structure is indeed rather complicated \cite{Nilles:2020nnc, Nilles:2020kgo, Nilles:2020tdp}.}

This paper is organized as follows.
In section 2, we introduce modular symmetry.
In section 3, we review the zero-mode profiles of magnetized torus.
We show how the wave functions and Yukawa couplings transform under the modular group.
In section 4, we study modular transformation of the Yukawa term.
We then investigate the modular flavor symmetry as the modular subgroup, 
under which the Yukawa term is invariant.
The group structure of modular flavor symmetry is also analyzed. 
In section 5 we consider modular transformation and flavor symmetry simultaneously.
We will show that they are noncommutative and they form a larger flavor group.
Section 6 is devoted to the conclusion.

\section{Modular Symmetry}

In this section, we introduce modular symmetry \cite{Feruglio:2017spp} 
and develop our notation.

The action of chiral superfields is determined by two functions:
K\"ahler potential $K$ and superpotential $W$.
Using these two functions, the action is given by
\begin{align}
\mathcal{S} = \int d^4 x d^2\theta d^2\bar \theta K(\Phi^i , \bar \Phi^i, \tau, \bar \tau)
+ \int d^4 x d^2\theta W(\Phi^i, \tau)  + (h.c.),
\end{align}
where $\Phi^i$ denotes a chiral superfield and $\tau$ is a complex parameter, i.e., modulus.
We assume $W$ is a holomorphic function of $\tau$ and $\Phi^i$, and $K$ is real.

Modular symmetry is the invariance of the action under modular transformation.
Let $\gamma$ be an element of $SL(2,\mathbb{Z})$.
Modular transformation of $\tau$ under $\gamma$ is given by
\begin{align}
\gamma~:~\tau \longmapsto \frac{a\tau+b}{c\tau+d},
\label{tra:mod_tau}
\end{align}
where $a,b,c,d$ are integers satisfying $ad-bc =1$.
Since the actions of $\gamma$ and $-\gamma$ are the same,
the modular transformation group $\Gamma$ is isomorphic to $PSL(2,\mathbb{Z}) = SL(2,\mathbb{Z})/\mathbb{Z}_2$.
The modular group is generated by two generators,
\begin{align}
S : \tau \longmapsto -\frac 1 \tau,
~~
T : \tau \longmapsto \tau +1,
\end{align}
and they correspond to the $SL(2,\mathbb{Z})$ elements as
\begin{align}
S = 
\begin{pmatrix}
  0 & 1 \\
-1 & 0
\end{pmatrix},
~~
T =
\begin{pmatrix}
1 & 1 \\
0 & 1
\end{pmatrix}.
\end{align}
Thus, modular invariance is equivalent to invariance under these two generators.

To construct modular invariant action, 
we introduce a holomorphic function known as modular form.
Modular forms are characterized by two parameters: weight $k$ and level $N$.
The modular group of level $N$ is a subgroup of the modular group given by
\begin{align}
\Gamma(N) = \left\{
\begin{pmatrix}
a & b\\
c & d
\end{pmatrix}
\in \Gamma
\;\middle|\;
a=d = 1~{\rm and}~b =c =0~{\rm mod}~ N
\right\},
\end{align}
and modular forms $f$ of weight $k$ and level $N$  are holomorphic functions of $\tau$,
which transform as
\begin{align}
f\left( \frac{a \tau+b}{c\tau+d} \right) = (c \tau +d)^k f(\tau),
\end{align}
under $
\begin{pmatrix}
a & b\\
c & d
\end{pmatrix}
\in 
\Gamma(N)$.
Let $f_1(\tau)$ and $f_2(\tau)$ be modular forms of weight $k$ and level $N$; then, $f_1(\tau) +f_2(\tau)$ is also a modular form of weight $k$ and level $N$.
Hence, the set of the modular forms of weight $k$ and level $N$ forms a vector space.
This space is denoted by $Mod_k^{(N)}$.
If $f(\tau)$ is a modular form of weight $k$ and level $N$, $f(\gamma\tau)$ is also a modular form of weight $k$ and level $N$.
This relation holds even if $\gamma \notin \Gamma(N)$.
Hence modular transformation of the modular forms can be written as
\begin{align}
f_i (\tau) \rightarrow (c \tau +d)^k \rho_{ij} f_j(\tau),
\end{align}
where $f_i$ is the basis of $Mod_k^{(N)}$, and $\rho$ is a unitary matrix.
$\rho$ is a representation of $\Gamma_N= \Gamma/\Gamma(N)$ since $\Gamma(N)$ trivially act on $Mod_k^{(N)}$. 
Modular forms are classified by the irreducible representations of $\Gamma_N$.
$\Gamma_N$ is a non-Abelian finite group if $N\leq 5$: $\Gamma_2 = S_3 , \Gamma_3 = A_4, \Gamma_4 = S_4, \Gamma_5 = A_5$ (and $\Gamma_1$ is a trivial group) \cite{deAdelhartToorop:2011re}.
The above non-Abelian groups have been used for non-Abelian flavor symmetries, and this is why modular symmetry is attractive for
particle phenomenology.

To construct modular invariant action,
we need modular transformations for chiral superfields.
We assume that each chiral superfield $\Phi^i$ is a modular form of weight $k_i$ and level $N$,
which transforms as 
\begin{align}
\Phi^i \rightarrow (c\tau+d)^{k_i} \rho_{k_i,ij}\Phi^j
\label{eq:chiral_mod}
\end{align}
under the modular group.
A modular invariant K\"ahler potential is given by
\begin{align}
K = \sum_i \frac {\Phi^i \bar \Phi^i} {(\tau - \bar \tau)^{-k_i}},
\end{align}
where $\im \tau$ transforms as $\im \tau \rightarrow 
|c\tau+d|^{-2} \im \tau$
under the modular group and it cancels the prefactor of \eqref{eq:chiral_mod}.
This form of the K\"ahler potential is obtained from dimensional reduction of superstring effective theory.
Construction of the modular invariant superpotential is more complicated.
We expand the superpotential $W$ as 
\begin{align}
W = \sum Y_{i_1 i_2 ... i_n}(\tau) \Phi^{i_1} \Phi^{i_2} ... \Phi^{i_n}.
\end{align}
We assume the coupling constant
$Y_{i_1 i_2 ... i_n}(\tau)$ is a modular form.
The modular invariant superpotential is realized if
the weight of $Y_{i_1 i_2 ... i_n}(\tau)$ is equal to $-k_{i_1} - k_{i_2} -... - k_{i_n}$,
and $\rho_{k_{i_{1}}} \otimes \rho_{k_{i_{2}}} \otimes ... \otimes \rho_{k_{i_{n}}} \otimes \rho_Y$ 
has the trivial singlet, where $\rho_Y$ is a representation of $Y$.

From a supergravity perspective, $\tau$ is a vacuum expectation value of the modulus field $U$ rather than a parameter, and
the superpotential is coupled to the K\"ahler potential.
The K\"ahler potential should include the kinetic term of $U$.
It is given by \cite{Dixon:1989fj}
\begin{align}
K_0 = - h \log (U +\bar U),
\end{align}
and $U$ is related to $\tau$ as $\tau = -i \braket{U}$.
The modular invariant condition is changed to \cite{Feruglio:2017spp}
\begin{align}
k_Y = -k_{i_1} -k_{i_2}- ...-k_{i_n} - h.
\label{eq:weight_con_SUGRA}
\end{align}

In the next section, we consider magnetized torus model.
In the following analysis we use canonically normalized chiral fields, 
and consider physical Yukawa couplings rather than holomorphic couplings.
The physical Yukawa couplings are no longer holomorphic function of the modulus, and 
their non-holomorphic part reflects the effects of K\"ahler potential.\footnote{
While our analysis is limited to global supersymmetry, 
the effect of the modular transformation of the tree-level K\"ahler potential for the complex structure moduli 
as well as for the matter field in Eq.~\eqref{eq:chiral_mod} can be identified with
the non-holomorphic part of the modular transformation of the physical Yukawa couplings 
via dimensional reduction of the ten-dimensional Yang-Mills theory with local 
supersymmetry \cite{Abe:2012ya} 
(see Eq. \eqref{eq:S-tensor}).
}
As we will see later, 
the modular invariance of the kinetic term (K\"ahler potential) of the matter fields is trivial 
as long as canonically normalized fields are used,
while they are not modular forms.
The modular invariance of the low-energy effective theory is investigated 
from the Yukawa interaction term (superpotential).

\section{Modular transformation in SYM theory on torus}

Let $\tau$ is a complex number satisfying $\im \tau >0$.
A lattice $L$ generated by $(1, \tau)$ is defined by
\begin{align}
L=\{n + m \tau \in \mathbb{C}~|~ \forall n, \forall m\in \mathbb{Z} \}
\nonumber
\end{align}
A torus is defined by $\mathbb{C}/L$.
Since the lattices generated by $(1,\tau)$ and $(a\tau+b, c\tau+d)$ are equivalent if $ad-bc =1$,
the modular group is symmetry of a torus.
$\tau$ is interpreted as the complex structure of a torus.
Thus, the natural origin of modular symmetric theories is a higher-dimensional theory compactified on a torus or its orbifold.
Indeed it is shown that effective action of the heterotic orbifolds is modular invariant \cite{Ferrara:1989bc}.
In this paper, we study modular invariance of six-dimensional SYM with $SU(N)$ compactified on a two-dimensional torus.
This model is known as magnetized torus, and it is the low-energy effective theory of type IIB superstring \cite{Blumenhagen:2000wh}.
Turning on background magnetic fluxes on the torus, the gauge group is broken to the direct product of its subgroup: $SU(N) \rightarrow SU(N_1) \times ... \times SU(N_\ell)$.
We assume $N = N_1 + \cdots +N_\ell$ in this paper, i.e., the Abelian Wilson line.
Such backgrounds break not only the gauge group
but also higher-dimensional supersymmetry,
and four-dimensional $N=1$ super Yang-Mills theory is realized as effective theory.
This property is certainly attractive for phenomenological purpose.
This model might be the origin of the Standard Model \cite{Abe:2008sx, Abe:2015yva, Marchesano:2004xz}.

To obtain the effective theory, we calculate mode expansion of bulk fields.
four-dimensional chiral superfields originate from the off-diagonal components of the gauginos.
After breaking the gauge group, they become bifundamental matter fields $\Phi_{ij}$, 
which transform as $(N_i, \bar N_j)$ under $SU(N_i) \times SU(N_j)$.
We briefly review the derivation of the zero-mode wave function of the $\Phi_{ij}$.
We consider the equation of motion for the fermionic component of $\Phi_{ij}$.
Wave functions of its scalar component are the same as those of the fermion unless four-dimensional supersymmetry is broken.
We also review modular transformation of the zero-modes and Yukawa couplings
\cite{Kobayashi:2016ovu, Kobayashi:2018bff, Kobayashi:2017dyu, Kobayashi:2018rad}.

The six-dimensional fields $\Phi$ are expanded by wave functions on the compact space,
\begin{align}\label{eq:6dim}
\Phi = \sum_n \phi_n(x) \psi_n(z,\bar z).
\end{align}
We concentrate on the zero-mode wave functions since we investigate modular invariance of low-energy effective theory.
The zero-mode equation for the fermionic components of $\Phi_{ij}$ is written as
\begin{align}
i \slashed D \psi = 
i\begin{pmatrix}
0 &  D^\dagger \\
D & 0 \\
\end{pmatrix}
\psi
=
\frac i{\pi R}
\begin{pmatrix}
0 &  \partial - \frac{\pi (m_i -m_j) }{2 \im \tau}(\bar z +\bar \zeta) \\
\bar \partial + \frac{\pi (m_i -m_j) }{2 \im \tau}(z +\zeta) & 0\\
\end{pmatrix}
\begin{pmatrix}
\psi_+(z,\tau)\\
\psi_-(z,\tau)
\label{eq:Dirac_eq}
\end{pmatrix}
=0,
\end{align}
where $z$ is the complex coordinate of the torus,
$\zeta$ is the Wilson line,
and $\partial$ is the partial derivative in terms of $z$.
$m_i, m_j$ are integer magnetic fluxes,
which are given by 
\begin{align}
F_{z\bar z} =\frac{\pi i}{\im \tau}
\begin{pmatrix}
m_1 1_{N_1 \times N_1 } &&\\
& \ddots \\
&&m_\ell 1_{N_ \ell \times N_\ell}
\end{pmatrix}.
\end{align}
The boundary conditions for the wave functions depend on the value of the magnetic flux.
They are summarized as the following two equations,
\begin{align}
\psi(z+1) &= \exp\left( i \frac{\pi M}{\im \tau} \im (z+\zeta) \right)\psi(z), 
\label{eq:boundary1}
\\
\psi(z+\tau) &= \exp\left(i \frac{\pi M}{\im \tau} \im \bar \tau (z+\zeta) \right)\psi(z),
\label{eq:boundary2}
\end{align}
where $M = m_i - m_j$.
The solutions of the Dirac equation are given by
\begin{align}
\psi^{j,M}_+(z,\tau) = \mathcal{N} e^{\pi i M (z+\zeta) {\rm Im} (z+\zeta)/{\rm Im} \tau} \TWC{\frac j M}{0} {M(z+\zeta)}{M\tau},
\label{eq:psi_positive}
\end{align}
for positive $M$, and 
\begin{align}
\psi^{j,M}_-(z,\tau) = \mathcal{N} e^{\pi i M (\bar z+ \bar \zeta) {\rm Im}(\bar z+ \bar \zeta)/{\rm Im} \bar \tau} \TWC{\frac j M}{0} {M(\bar z +\bar \zeta) }{M\bar\tau},
\label{eq:psi_negative}
\end{align}
for negative $M$.
$j$ runs from $0$ to $|M|-1$ for the both cases.
Thus we have $|m_i-m_j|$ replicas of zero-modes for each $\Phi_{ij}$.
This is the origin of the generations of the quarks and the leptons \cite{Abe:2008sx, Abe:2015yva, Marchesano:2004xz}
.
$\TWC{\alpha}{\beta}{z}{\tau}$ is the Jacobi theta function:
\begin{align}
\TWC{\alpha}{\beta}{z}{\tau}
= \sum_{n\in \mathbb{N}} e^{\pi i (n+\alpha)^2 \tau} e^{2\pi i (n+\alpha)(z+\beta)}. 
\end{align}
Since the Jacobi theta function can not be well defined if $\im \tau \leq 0$, $\psi_+$ have the normalizable solutions only when $M >0$, 
and $\psi_-$ becomes normalizable only when $M<0$.
Hence chiral theory is realized.
Using the area of the torus $\mathcal{A}$,
a normalization factor $\mathcal{N}$ is calculated as
\begin{align}
\mathcal{N} = \left(\frac{2 |M| \im\tau}{\mathcal{A}^2} \right)^{\frac 1 4}.
\end{align}

The action of $\gamma$ on the zero-mode wave function is defined as
\begin{align} \label{eq:mtrwf}
\psi (z,\tau) \rightarrow \psi' = \psi\left(
\frac z {c \tau+d}, \frac{a\tau+b}{c\tau+d}
\right),
\end{align}
where $ad-bc =1$ \cite{Mumford:1983}.
It is easily checked that antiholomorphic part of $\psi_+$ and holomorphic part of $\psi_-$ are not changed by the modular transformation. 
Since the Dirac operator includes only $\bar \partial$ for $\psi_+$ and $\partial$ for $\psi_-$,
the wave function $\psi'$ also satisfies the original zero-mode Dirac equation $D\psi' = 0$ for any $\gamma \in SL(2,Z)$. 
Indeed substituting $\psi'_+$ to $\psi_+$ in \eqref{eq:Dirac_eq}, we obtain
\begin{align}
D\psi'_+ = D\psi\left(
\frac z {c \tau+d}, \frac{a\tau+b}{c\tau+d}
\right)
=\left(M \pi i \frac {z+\zeta} {c\tau+d} \frac{-(c\tau+d)}{2i \im \tau} 
+ \frac{\pi M}{2 \im \tau}(z +\zeta) \right) \psi_+ =0.
\end{align}
The same relation holds for $\psi_-$.
However, the boundary conditions \eqref{eq:boundary1} and \eqref{eq:boundary2} are not always satisfied.
Define a new holomorphic function $f(z)$ by
\begin{align}
\psi^{j,M}\left(
\frac z {c \tau+d}, \frac{a\tau+b}{c\tau+d}
\right)
&=
\mathcal{N} e^{i \frac{M\pi}{2\im \tau} \im (z+\zeta)^2} f(z).
\end{align}
The boundary conditions for the wave function are reinterpreted to the conditions for $f(z)$.
Equations \eqref{eq:boundary1} and \eqref{eq:boundary2} are equivalent to
\begin{align}
f(z+ a\tau + b) &= e^{-\pi i a^2 M \re \tau} e^{-2 \pi i a M \re (z + \zeta)} f(z),
\nonumber
\\
f(z + c\tau + d) &= e^{-\pi i c^2 M \re \tau}e^{-2 \pi i c M \re (z+\zeta)} f(z).
\end{align}
On the other hand, the zero-mode wave functions \eqref{eq:psi_positive} and \eqref{eq:psi_negative} imply that
\begin{align}
f(z+a\tau+b)
&=
e^{-2 M \pi i a \re (z + \zeta)} e^{-M\pi i a^2 \re \tau -M \pi i ab}
f(z),
\nonumber
\\
f(z+c\tau+d) 
&=
e^{-2 M\pi i c \re (z + \zeta) -M\pi i c^2 \re \tau -M\pi i cd } 
f(z).
\end{align}
Thus the boundary conditions are satisfied only when $M c d$ and $M ab$ are even.
When $M$ is even, these conditions are satisfied for all $a, b, c, d$, 
and the action of $\gamma$ is well defined.
When $M$ is odd, the action of $\gamma$ is not consistent 
with the boundary conditions if $ab$ or $cd$ is odd.
For odd $M$, however, it is found that a subgroup such that $ab$ and $cd$ are even
is consistent with the boundary conditions.
This subgroup is called $\Gamma_{1,2}$ \cite{Mumford:1983}:
\begin{align}
\Gamma_{1,2} = \left\{
\begin{pmatrix}
a & b\\
c & d
\end{pmatrix}
\in SL(2,Z)
\;\middle|\;
ab, cd \in 2\mathbb{Z}
\right\}.
\end{align}
Now we can define modular transformation (or transformation under $\Gamma_{1,2} $) of the matter fields.
We summarize their results.
Let $M$ be a positive integer.
Then, the transformation of the wave function under $S$ is given by
\begin{align}
\psi^{j,M} (- z/\tau,-1/\tau) 
&= \left(\frac{2 M {\rm Im}\frac{-1}{\tau} }{\mathcal{A}^2} \right)^{\frac 1 4}
e^{\pi i M \frac {z+\zeta} \tau {\rm Im} {(z+\zeta)} \bar \tau/{\rm Im} \tau}
\TWC{\frac j M}{0}{- M{(z+\zeta)}/\tau}{-M/\tau}
\nonumber
\\
&=\frac {e^{-\frac{\pi i}4}} {\sqrt{M}} \left(\frac \tau {|\tau|}\right)^{\frac 1 2}  \sum_k e^{2\pi i \frac{jk}M} \psi^{k,M}(z,\tau).
\label{eq:mod_tra_S}
\end{align}
In the second row, we use modular transformation of Jacobi theta function
\begin{align}
\vartheta
\begin{bmatrix}
-\beta\\
\alpha
\end{bmatrix}
(z,\tau)
&= (-\tau)^{-1/2}
e^{-\pi i  \frac{z^2}{\tau}}
\vartheta
\begin{bmatrix}
\alpha\\
\beta
\end{bmatrix} 
\left(\frac{-z}{\tau},\frac{-1}{\tau}\right),
\label{eq:S-trans_Jacobi}
\end{align}
and the Poisson resummation formula
\begin{align}
 \TWC{0}{\frac j N }{\nu}{\tau/N} =  \sum_{k=0,..,N-1} e^{2\pi i \frac{jk}N}
 \TWC{\frac k N }{0}{N\nu}{N\tau}.
\label{eq:chi-psi}
\end{align}
If $M$ is even, 
the modular transformation of the wave function under $T$ is 
given as 
\begin{align}
\psi^{j,M}(z, \tau+1) &= e^{\pi i\frac{j^2}M}  \psi^{j,M}(z,\tau).
\label{eq:mod_tra_T}
\end{align}
Since $\Gamma$ is generated by $S$ and $T$, 
we obtain the modular transformation of the chiral zero-modes for even $M$.
If $M$ is odd,
as shown before, we consider modular transformation of the subgroup $\Gamma_{1,2}$. 
Since all the elements of $\Gamma_{1,2}$ are generated 
by $S$ and $T^2$, 
we consider the modular transformation of the zero-modes under $T^2$,
which is calculated as
\begin{align}
\psi^{j,M}(z, \tau+2)  = e^{2\pi i \frac{j^2} M} \psi^{j,M}.
\label{eq:mod_tra_T^2}
\end{align}
In the case of negative $M$, 
modular transformation is given as 
the complex conjugate of the one for $\psi^{j,|M|}_+(z)$
since $\psi^{j,M}_-(z)$ is the complex conjugate of $\psi^{j,|M|}_+(z)$.

We introduce a matrix representation for $S$ and $T$ as
\begin{align}
&\psi^{j,M}\left(\frac z{\tau}, -\frac{1}{\tau} \right)=
e^{-\frac{\pi i}4}  
\left(\frac{\tau}{|\tau|} \right)^{1/2}
\rho_M(S)_{jk} \psi^{k,M}(z,\tau),
\\
&\psi^{j,M}(z,\tau+1)=
\rho_M(T)_{jk} \psi^{k,M}(z,\tau),
\end{align}
for positive and even $M$.
$\rho_M(S)$ and $\rho_M(T)$ are a matrix representation 
for the $M$-component vector of the chiral zero-modes, 
which are denoted by
\begin{align}
&\rho_M(S) = \frac1{\sqrt{M}}
\begin{pmatrix}
1  & 1 & \cdots & 1\\
1  & \sigma & \cdots & \sigma^{M-1}\\
\vdots &  & \ddots & \vdots  \\
1  & \sigma^{M-1} & \cdots & \sigma
\end{pmatrix},
\label{eq:S-matrix}
\\
&\rho_M(T) =
\begin{pmatrix}
1  & 0 & \cdots & 0\\
0 & e^{\pi i \frac{1}M} & \cdots & 0\\
\vdots &  & \ddots & \vdots  \\
0  & 0 & \cdots & e^{\pi i \frac{(M-1)^2}M}
\end{pmatrix},
\label{eq:T-matrix}
\end{align}
where $\sigma= e^{\frac{2\pi i}{M}}$.
$\rho_M(S)$ and $\rho_M(T)$ are noncommutative with each other, and they generate a non-Abelian finite group.
If $M$ is odd, we consider $T^2$ instead of $T$ and its matrix representation is given as 
\begin{align}
\psi^{j,M}(z,\tau+2)=
\rho_M(T^2)_{jk} \psi^{k,M}(z,\tau).
\end{align}
The matrix representation for negative $M$ 
is given as the complex conjugate of the one for positive $M$: 
\begin{align}
\rho_M(S)= (\rho_{|M|}(S))^*, \quad  
\rho_M(T)= (\rho_{|M|}(T))^*.
\end{align}
We note that the modular transformation given by $\rho_M(S)$ and $\rho_M(T)$
is a unitary transformation among the zero-mode wave functions.

We consider the modular transformation of the Yukawa couplings.
Four-dimensional effective couplings are calculated by overlap integrals 
among the zero-mode wave functions.
Yukawa couplings of magnetized torus are given by \cite{Cremades:2004wa}
\begin{align}
Y_{ij\bar k} = \int_{T^2} d z d\bar z \psi^{i,M_1} \psi^{j.M_2} (\psi^{k,|M_3|})^*
\label{eq:Yukawa_over}
\end{align}
where we assume that $M_1$ and $M_2$ are positive and $M_3$ is negative for definiteness.
$M_1 + M_2 +M_3 =0$ since $M_i = m_j -m_k$.
Substituting the zero-mode wave functions in \eqref{eq:Yukawa_over},
we obtain Yukawa couplings: 
\begin{align}
Y_{ijk}(\tau) =& \left( \frac{2 {\rm Im} \tau}{\mathcal{A}^2}\right)^{1/4} \left|\frac{M_1 M_2}{M_3} \right|^{1/4} 
e^{\frac{\pi i}{\im \tau}\sum_i M_i \zeta_i \im \zeta_i}
\nonumber
\\
&\sum_{m\in \mathbb{Z}_{M_3}}
\delta_{k,i+j+M_1 m}
\vartheta
\begin{bmatrix}
\frac{M_2 i - M_1 j +M_1 M_2 m}{-M_1 M_2 M_3}\\
0
\end{bmatrix}
(\tilde \zeta ,|M_1 M_2 M_3|\tau),
\label{eq:Yukawa}
\end{align}
where the Kronecker delta is defined modulo $M_3$, which means 
$\delta_{k,i+j+M_1 m} =1$ if and only if $k=i+j+M_1 m$ mod $M_3$.
The index $i$ runs from 0 to $M_1-1$, $j$ runs from $0$ to $M_2-1$, and $k$ runs from $0$ to $|M_3|-1$.
$\zeta_i$ is the Wilson line corresponding to $M_i$, and $\tilde \zeta$ is given by $\tilde \zeta= M_1 M_2 (\zeta_1- \zeta_2)$.
From Eq.~\eqref{eq:Yukawa}, 
the action of $S$ and $T$ on the Yukawa couplings can be read off as
\begin{align}
Y_{ijk}\left(-\frac1\tau\right) =& \left( \frac{2 {\rm Im} \tau}{|\tau|^2 \mathcal{A}^2}\right)^{1/4} \left|\frac{M_1 M_2}{M_3} \right|^{1/4}
e^{\frac{\pi i}{\im \tau}\sum_i M_i \frac{\zeta_i}{\tau} \im \zeta_i \bar\tau}
\nonumber
\\
&\sum_{m\in \mathbb{Z}_{M_3}}
\delta_{k,i+j+M_1 m}
\vartheta
\begin{bmatrix}
\frac{M_2 i - M_1 j +M_1 M_2 m}{-M_1 M_2 M_3}\\
0
\end{bmatrix}
\left(\frac{\tilde \zeta}{\tau},-\frac{|M_1 M_2 M_3|}{\tau}\right),
\nonumber
\\
= &\left( \frac{2 {\rm Im} \tau}{|\tau|^2 \mathcal{A}^2}\right)^{1/4} \left|\frac{M_1 M_2}{M_3} \right|^{1/4} 
\left(
\frac {-i \tau}{|M_1 M_2 M_3|}
\right)^{1/2}
e^{\frac{\pi i}{\im \tau}\sum_i M_i \zeta_i \im \zeta_i}
\nonumber
\\
&
\sum_{m\in \mathbb{Z}_{M_3}}
\delta_{k,i+j+M_1 m}
\sum_{\ell=0,...,|M_1 M_2 M_3|-1}
e^{2\pi i \frac{(M_2 i - M_1 j +M_1 M_2 m)\ell}{|M_1 M_2 M_3|}}
\TWC{\frac{\ell}{|M_1 M_2 M_3|}}{0}{\tilde \zeta}{|M_1 M_2 M_3|\tau},
\label{eq:Y_S_general}
\end{align}
and
\begin{align}
Y_{ijk}\left(\tau +1 \right) 
&= \left( \frac{2 {\rm Im} \tau}{\mathcal{A}^2}\right)^{1/4} \left|\frac{M_1 M_2}{M_3} \right|^{1/4} 
e^{\frac{\pi i}{\im \tau}\sum_i M_i \zeta_i \im \zeta_i}
\sum_{m\in \mathbb{Z}_{M_3}}
\delta_{k,i+j+M_1 m}
\nonumber
\\
&~~~~~~~
e^{\pi i \frac{(M_2 i -M_1 j + M_1 M_2 m)^2}{|M_1 M_2 M_3|}}
\TWC{\frac{M_2 i - M_1 j +M_1 M_2 m}{-M_1 M_2 M_3}}{0}{\tilde \zeta}{\frac{\tau}{|M_1 M_2 M_3|}},
\label{eq:Y_T_general}
\end{align}
where we use the fact that $M_1 M_2 M_3$ is even for $|M_3| = M_1 +M_2$.
When the greatest common divisor of $M_1, M_2$, and $|M_3|$ is 1, 
the Yukawa couplings can be written in a simpler form:
\begin{align}
Y_{ijk}(\tau) =  \left( \frac{2 {\rm Im} \tau}{\mathcal{A}^2}\right)^{1/4} \left|\frac{M_1 M_2}{M_3} \right|^{1/4} 
e^{\frac{\pi i}{\im \tau}\sum_i M_i \zeta_i \im \zeta_i}
\TWC{\frac i{M_1} + \frac j{M_2} +\frac k{M_3} }{0}{\tilde \zeta}{|M_1 M_2 M_3|\tau}.
\label{eq:Yukawa_gcd1}
\end{align}
In this case, 
modular transformation is given by
\begin{align}
Y_{ijk}\left(-\frac1\tau\right)
&=  \left( \frac{2 {\rm Im} \tau}{\mathcal{A}^2|\tau|^2}\right)^{1/4} \left|\frac{M_1 M_2}{M_3} \right|^{1/4} e^{\frac{\pi i}{\im \tau}\sum_i M_i \frac{\zeta_i}{\tau} \im \zeta_i \bar\tau}
\TWC{\frac i{M_1} + \frac j{M_2} +\frac k{M_3} }{0}{\frac{\tilde \zeta}{\tau}}{-\frac{|M_1 M_2 M_3|}{\tau}}
\nonumber
\\
&= \left( \frac{2 {\rm Im} \tau}{\mathcal{A}^2|\tau|^2}\right)^{1/4} \left|\frac{M_1 M_2}{M_3} \right|^{1/4} \left(\frac{\tau}{|M_1M_2 M_3|}\right)^{1/2} e^{-\frac {\pi i}4}
e^{\frac{\pi i}{\im \tau}\sum_i M_i \zeta_i \im \zeta_i}
\nonumber
\\
&~~~~~~~
\sum_{\ell=0, ... , |M_1 M_2 M_3|-1} e^{2\pi i \frac{-i M_2 M_3 - j M_3 M_1 - k M_1 M_2}{|M_1 M_2 M_3|} \ell}
\TWC{\frac{\ell}{ |M_1 M_2 M_3|}}{0} {\tilde \zeta}{|M_1 M_2 M_3|\tau},
\nonumber
\\
&=   \left(\frac{\tau}{|\tau|}\right)^{1/2} e^{-\frac {\pi i}4}
\sum_{\ell=0, ... , |M_1 M_2 M_3|-1}
\frac{1}{\sqrt{M_1 M_2 M_3}} e^{2\pi i \frac{-i M_2 M_3 - j M_3 M_1 - k M_1 M_2}{|M_1 M_2 M_3|}\frac{i'}{M_1}+\frac{j'}{M_2}+\frac{k'}{M_3} }
Y_{i' j' k'},
\label{eq:S-Yukawa_gcd1}
\end{align}
and
\begin{align}
Y_{ijk}\left(\tau+1\right)
&= 
\left( \frac{2 {\rm Im} \tau}{\mathcal{A}^2}\right)^{1/4} 
\left|\frac{M_1 M_2}{M_3} \right|^{1/4} 
e^{\frac{\pi i}{\im \tau}\sum_i M_i \zeta_i \im \zeta_i}
\TWC{\frac i{M_1} + \frac j{M_2} +\frac k{M_3} }{0}{\tilde \zeta}{|M_1 M_2 M_3|\tau +|M_1 M_2 M_3|}
\nonumber
\\
&=e^{\pi i \left(\frac{(-i M_2 M_3 - j M_3 M_1 - k M_1 M_2)^2}{|M_1 M_2 M_3|} \right)} Y_{ijk}(\tau)
\label{eq:T-Yukawa_gcd1}
\end{align}
Therefore the Yukawa couplings form a representation of the modular group.

It is shown here that 
the modular transformation of the Yukawa couplings 
is given as a linear combination of the original Yukawa couplings. 
This is because the Yukawa couplings are given by the overlap integral of the zero-modes, 
so the modular transformation of the Yukawa couplings 
is given by a tensor product of the modular transformation of each zero-mode.
Thus, they form a representation of the modular group.
In fact the modular transformation of the Yukawa couplings 
given in Eqs.~\eqref{eq:Y_S_general} and \eqref{eq:Y_T_general} 
is equivalent to the tensor representation
\begin{align}
\label{eq:S-tensor}
Y_{ijk}\left(-\frac1\tau\right) 
&=e^{-\frac{\pi i}{4}} 
\left(\frac{\tau}{|\tau|}\right)^{1/2}
\rho_{M_1}(S)_{ii'}
\rho_{M_2}(S)_{jj'}
(\rho_{|M_3|}(S)_{kk'})^*  Y_{i'j'k'}(\tau), 
\\
Y_{ijk}(\tau+1)
&=\rho_{M_1}(T)_{ii'}\rho_{M_2}(T)_{jj'}(\rho_{|M_3|}(T)_{kk'})^*  Y_{i'j'k'}(\tau),
\end{align}
which will be used for the analysis of the modular invariance of the Yukawa term 
in the next section.

In what follows
we ignore overall $U(1)$ phases such as $e^{-\frac{\pi i}{4}}$ 
which appear in the modular transformations 
for the matter fields and the Yukawa couplings,  
since it can always be rotated away by field redefinition.

\section{Modular Flavor Symmetry on Magnetized Torus}

\subsection{Local supersymmetry and the Yukawa interaction}

The effective theory of the magnetized torus is consistent with local supersymmetry if the Wilson line vanishes \cite{Abe:2012ya}.%
\footnote{For nonvanishing Wilson line, the situation is more complicated.
It is unclear how to split the interaction term into a holomorphic part and real part.}
The physical Yukawa coupling is given in supergravity as
\begin{align}
Y_{ijk} = e^{K_0/2} (K_{i\bar i} K_{j \bar j} K_{k \bar k})^{-1/2} y_{ijk},
\end{align}
where $K_0$ is the K\"ahler potential of moduli fileds, $K_{i\bar i}$ is that of the matter fields,
and $y_{ijk}$ is the holomorphic Yukawa coupling.
The effective action of type IIB superstring implies
\begin{align}
K_0 &\sim -\ln(U + \bar U) + ...
\end{align}
$U$ is the complex structure moduli field: $i \braket{U} = \tau $.
We omit the K\"ahler potential of K\"ahler modulus $T$ and the dilaton $S$ since it is irrelevant to the modular symmetry.
The K\"ahler potential in terms of the chiral superfields and the superpotential is given by 
\begin{align}
K & \sim \sum_{j, M} \frac{\tilde \phi^{j,M} \bar {\tilde \phi}^{j,M}}{(U+\bar U)^{1/2}},
\nonumber
\\
W  & \sim  \left|\frac{M_1 M_2}{M_3} \right|^{1/4} 
\TWC{\frac i{M_1} + \frac j{M_2} +\frac k{M_3} }{0}{0}{ |M_1 M_2 M_3| i U} \phi^{j, M_1} \phi^{k, M_2} \phi^{\ell,|M_3|},
\end{align}
where we omit the $S$ and $T$ dependent terms too.
The modular weights of the chiral superfields are $-1/2$.
The modular transformation of the Jacobi theta function \eqref{eq:S-trans_Jacobi} implies that the weight of the holomorphic Yukawa couplings is $1/2$. 
Thus they satisfy the modular invariant condition \eqref{eq:weight_con_SUGRA}.

We investigate the modular symmetry of the Yukawa term:
\begin{align}
Y_{j k \ell } 
\phi^{j,M_1} \phi^{k,M_2} \phi^{\ell,|M_3|}
\nonumber
\end{align}
where $\phi^{j, M_k}$ denotes the four-dimensional chiral field in Eq.~\eqref{eq:6dim}.
$\phi^{j,M}$ is a canonically normalized chiral superfield and it corresponds to $\tilde \phi^{j,M}$ as $\phi^{j,M} \propto \im \tau^{-1/4} \tilde \phi^{j,M}$.

Modular transformation of the four-dimensional fields $\phi$
should coincide with that of the wave functions on the compact space,
since the six-dimensional fields should be invariant under the modular group.%
\footnote{If the modular group acts on the six-dimensional fields nontrivially,
their representations might be different, but we ignore this possibility in this paper.}
This is the same as the flavor symmetry originating from extra dimensions \cite{Abe:2009vi}.
Thus the modular transformation for the four-dimensional fields is written as
\begin{align}
\tilde \phi^{j,M} \rightarrow (c\tau+d)^{-1/2} \rho_{M, jk} \tilde \phi^{k, M},
\end{align}
and the modular transformation of the canonically normalized chiral superfield is given by
\begin{align}
\phi^{j,M} \rightarrow \left(\frac{|c \tau +d |}{c\tau+d}\right)^{1/2} \rho_{M, jk} \phi^{k,M}.
\end{align}

Using the tensor representation 
we obtain the general modular transformation of the Yukawa term by $g\in \Gamma$ as
\begin{align}
Y_{j k \ell } 
\phi^{j,M_1} \phi^{k,M_2} \phi^{\ell,|M_3|}
&\xrightarrow{g}
\rho_{M_1, j j'} \rho_{M_2, k k'} \rho_{|M_3|, \ell \ell'}^* Y_{j'k'\ell'} 
\rho_{M_1, j j''} 
\phi^{j'',M_1} 
\rho_{M_2, k k''}\phi^{k'',M_2} 
 \rho_{|M_3|, \ell \ell''}^* \phi^{\ell'',|M_3|}
\nonumber
\\
&=\,
(\rho_{M_1}^T \rho_{M_1})_{j'' j'} (\rho_{M_2}^T \rho_{M_2})_{k'' k'} (\rho_{|M_3|}^\dagger \rho^*_{|M_3|})_{\ell'' \ell'}
Y_{j'k'\ell'} 
\phi^{j'',M_1} 
\phi^{k'',M_2} 
\phi^{\ell'',|M_3|}.
\label{eq:yt_tensor}
\end{align}
Here the overall phases are ignored.
We obtain the Yukawa invariant modular subgroup $\mathcal{M}$ by
\begin{align}
\mathcal{M} &= \left\{
g \in \Gamma
\;\middle|\; 
\tilde{\rho}_{M_1}(g)_{j j'} 
\tilde{\rho}_{M_2}(g)_{k k'}
\tilde{\rho}_{|M_3|}^*(g)_{\ell \ell'}
Y_{j' k' \ell'}
= Y_{jk\ell}
\right\}
\label{eq:rhoTrho},
\end{align}
where 
$\tilde{\rho}_M(g)$ is defined 
as $\tilde{\rho}_M(g) = \rho_M^T(g) \rho_M (g)$.
Hereafter we refer to the Yukawa invariant modular subgroup $\mathcal{M}$ as the modular flavor symmetry.

We show that 
the Yukawa invariant modular subgroup $\mathcal{M}$ has 
the following three independent elements of 
$S^2$, $T^N$  and $(ST^N)^2$, 
where $N$ is the least common multiple of the generation numbers of the corresponding zero-modes.
($T^N$ is well defined since $N$ is always even.)
The representations of $S^2$ and $T^N$ are written as
\begin{align}
\rho_M(S)^2 &= 
\begin{pmatrix}
1  & 0 & \cdots  & 0\\
0  & 0 & \cdots  & 1\\
\vdots & \vdots & \ddots & \vdots  \\
0  & 1 & \cdots &  0
\end{pmatrix},
\label{eq:S^2-matrix}
\\
\rho_M(T^N) &=
\begin{pmatrix}
1  & 0 & \cdots & 0\\
0 & e^{N \pi i \frac{1}M} & \cdots & 0\\
\vdots & \vdots & \ddots & \vdots  \\
0  & 0 & \cdots & e^{N \pi i \frac{(M-1)^2}M}
\end{pmatrix}
=
\begin{pmatrix}
1  & 0 & \cdots & 0\\
0 & (-1)^{N/M} & \cdots & 0\\
\vdots & \vdots & \ddots & \vdots  \\
0  & 0 & \cdots & (-1)^{N/M}
\end{pmatrix}.
\label{eq:T^N-matrix}
\end{align}
There are two cases for the matrix representations of $T^N$ and $(S T^N)^2$.
If $M$ is even and $N/M$ is odd, 
since $\rho_M(T^N)$ is not the identity, 
the $\rho_M((ST^N)^2)$ is given by
\begin{align}
&\rho_M((ST^N)^2)_{ij} =(-1)^{i-1} \delta^{(M)}_{i, -j - \frac M 2},
\label{eq:ST^N^2}
\end{align}
where the index runs from 0 to $M-1$ and the Kronecker delta is defined modulo $M$;
otherwise, $\rho_M(T^N)=1$ and $\rho_M((ST^N)^2) = \rho_M(S^2)$.
Through these matrices, we can check the invariance of the Yukawa term.
$S^2$ and $T^N$ invariance is obvious since 
\begin{align}
\rho_M(S^2)^T \rho_M(S^2) = \rho_M(T^N)^T \rho_M(T^N) = 1.
\end{align}
For $\rho_M((ST^N)^2)$,
if $M$ is even and $N/M$ is odd, 
substituting \eqref{eq:ST^N^2}, we find
\begin{align}
\rho_M((ST^N)^2)^T \rho_M((ST^N)^2)= (-1)^{i-1} \delta_{i, -j-\frac M 2} \delta (-1)^{k-1} \delta_{k,-j-\frac M 2} = \delta_{i,k}
\end{align}
Thus, the Yukawa term is $(ST^N)^2$ invariant too.

In the case of vanishing Wilson line, 
the modular symmetry is enhanced.
In this case we have $\mathbb{Z}_2$ parity symmetry~\cite{Abe:2009vi}:
\begin{align}
\phi^{j,M} =\phi^{M-j,M}.
\end{align}
Substituting the $\rho_M(S)$ into \eqref{eq:yt_tensor}, we find
\begin{align}
Y_{j k \ell } \phi^{j,M_1} \phi^{k,M_2} \phi^{\ell,|M_3|}
&\xrightarrow{S} 
(\rho_{M_1}(S))^2_{jj'} (\rho_{M_2}(S))^2_{kk'} (\rho^*_{|M_3|}(S))^2_{\ell \ell'} 
Y_{j'k'\ell'} \phi^{j,M_1} \phi^{k,M_2} \phi^{\ell,|M_3|}
\nonumber
\\
&=  
Y_{j k \ell} \phi^{M_1-j,M_1} \phi^{M_2-k,M_2} \phi^{|M_3| - \ell,|M_3|},
\notag \\
&=  
Y_{j k \ell } \phi^{j,M_1} \phi^{k,M_2} \phi^{\ell,|M_3|}
\label{eq:Y_S-trans}
\end{align}
in the second row, we use \eqref{eq:S^2-matrix}.
The Yukawa term is $S$ invariant.
Therefore, in the case of vanishing Wilson line, 
the Yukawa invariant modular subgroup $M$ 
has two independent generators of $S$ and $T^N$.
We will see that $S$ can be interpreted as a ``square root'' of the parity operator
in Section \ref{Sec:MEDS}.

\subsection{Modular flavor symmetry in three-generation model}

In this section we study a characteristic example of the three generations 
to illustrate the modular flavor symmetry.
Suppose that the gauge group $SU(N)$ is broken to three non-Abelian gauge groups, $SU(N_1) \times SU(N_2) \times SU(N_3)$, and integer magnetic fluxes of $m_1, m_2, m_3$ are turned on.
Let $M_1=M_2=3$ and $M_3=-6$.
In this case, there are two three-generation chiral zero-modes and one six-generation chiral zero-mode.

\subsubsection*{Model with Wilson line}

First we consider the case with nonvanishing Wilson line.
The wave functions for three-generation chiral zero-modes are given by
\begin{align}
&\psi^{j,3}=\mathcal{N} e^{\pi i 3 (z+\zeta) {\rm Im} (z+\zeta)/{\rm Im} \tau}
 \TWC{\frac j 3}{0} {3(z+\zeta)}{3\tau},
\end{align}
where $j=0,1,2$.
The modular transformations of these wave functions are given by \eqref{eq:mod_tra_S} and \eqref{eq:mod_tra_T}.
For $M=3$, the matrix representations are given by
\begin{align}
\rho_3(S)= \frac 1 {\sqrt{3}}
\begin{pmatrix}
1 & 1 &1\\
1 & \omega & \omega^2\\
1 & \omega^2 & \omega
\end{pmatrix},
~~
\rho_3(T^2)=
\begin{pmatrix}
1 & 0 & 0\\
0 & \omega & 0\\
0 & 0 & \omega
\end{pmatrix},
\label{eq:matrix_3-generation}
\end{align}
where $\omega = e^{\frac{2\pi i}3} $.
We study $T^2$ instead of $T$ since $M_{1,2}$ are odd.
For $M=|M_3|=6$, the matrix representations are given by 
\begin{align}
\rho_6(S)= \frac 1 {\sqrt{6}}
\begin{pmatrix}
1 & 1 &1 & 1 & 1 & 1\\
1 & \eta &\eta^2 & -1 & \eta^4 &\eta^5\\
1 & \eta^2 &\eta^4 & 1 & \eta^2 &\eta^4\\
1 & -1 &1 & -1 & 1 &-1\\
1 & \eta^4 &\eta^2 & 1 & \eta^4 &\eta^2\\
1 & \eta^5 &\eta^4 & -1 & \eta^2 &\eta^1\\
\end{pmatrix},
~~
\rho_6(T^2)=
\begin{pmatrix}
1 & 0 & 0 & 0 & 0& 0\\
0 & \eta & 0& 0& 0& 0\\
0 & 0 & \eta^2 & 0 & 0& 0\\
0 & 0 & 0 & -1 & 0& 0 \\
0 & 0 & 0 & 0 & \eta^4 & 0 \\
0 & 0 & 0 & 0 & 0& \eta \\
\end{pmatrix},
\label{eq:M=6_matrix}
\end{align}
where $\eta = e^{\frac{\pi i}{3}}$.
The Yukawa couplings $Y_{ijk}$ are classified into six values:
\begin{align}
&
Y_0 \equiv Y_{000} = Y_{112} = Y_{224},~~~
Y_1 \equiv Y_{101} = Y_{213} = Y_{025},~~~
Y_2 \equiv Y_{120} = Y_{202} = Y_{014},
\nonumber
\\
&
Y_3 \equiv Y_{221} = Y_{003} = Y_{115},~~~
Y_4 \equiv Y_{210} = Y_{022} = Y_{104},~~~
Y_5 \equiv Y_{011} = Y_{123} = Y_{205},
\end{align}
where $Y_j$ is given by
\begin{align}
Y _ j(\tau) = \left( \frac{3 {\rm Im} \tau}{\mathcal{A}^2}\right)^{1/4}
\left\{
 \TWC{\frac j {18} }{0}{\tilde\zeta}{54\tau} + \TWC{\frac {j+6}{18}}{0}{\tilde\zeta}{54\tau} +\TWC{\frac {j+12}{18}}{0}{\tilde\zeta}{54\tau}\right\}.
\end{align}
Other couplings are prohibited by the $\mathbb{Z}_3$ charge of $\Delta(27)$ flavor symmetry \cite{Abe:2009vi}.
A matrix representation of the modular transformation 
for the 6-component vector ($Y_i$) 
is defined as 
\begin{align}
Y_j \left(-\frac{1}{\tau} \right) =
\rho_Y(S)_{jk} Y_k(\tau), \quad 
Y_j (\tau+1) =\rho_Y(T)_{jk} Y_k(\tau).
\end{align}
In this basis, $\rho_Y$ is exactly the same as the one  
for the six-generation chiral zero-mode, i.e., $\rho_Y=\rho_6$.

The Yukawa invariant modular subgroup is generated by $S^2$, $T^6$ and $(ST^6)^2$.
These elements satisfy the following relations:
\begin{align}
\rho_M(S^2)^2 = \rho_M(T^6)^2 = \rho_M((ST^6)^2)^4 =1.
\end{align}
Thus, they correspond to $\mathbb{Z}_2$ and $\mathbb{Z}_4$ respectively.
$(ST^6)^2$ and $T^6$ are noncommutative, and these three elements generate a non-Abelian group.
This group has 16 elements.
and is found to be isomorphic 
to $\mathbb{Z}_2^{(S^2)}\times (\mathbb{Z}_4^{((T^6 S)^2)}\rtimes \mathbb{Z}_2^{(T^6)}) = \mathbb{Z}_2\times D_4$.
The irreducible decomposition of the chiral zero-modes is given by
\begin{align}
{\bf 3} &= {\bf 1}_{++}^+\oplus{\bf 1}_{++}^+\oplus {\bf 1}_{+-}^-,\\
{\bf 6} &= {\bf 2}^{+}\oplus{\bf 2}^{+}\oplus {\bf 2}^{-},
\end{align}
where the lower index of ${\bf 1}$ denotes the eigenvalues of $T^6$ and $(S T^6 )^2$,
and the upper index denotes the eigenvalue of the diagonal $\mathbb{Z}_2$.
Since $\mathbb{Z}_2$ and $D_4$ are real, irreducible decomposition of 
the Yukawa couplings is the same as that of $\psi^{j,6}$.
Table~\ref{tab:irre_336} summarizes the irreducible decomposition of each component.

\begin{table}
\begin{center}
\begin{tabular}{|c | c |} 
\hline
& 
Representation of $D_4 \times \mathbb{Z}_2$
\\
\hline \hline
$\psi^{j,3}$
& ${\bf 1}_{++}^+\oplus{\bf 1}_{++}^+\oplus {\bf 1}_{+-}^-$ 
\\
$\psi^{j,6}$
& ${\bf 2}^{+}\oplus{\bf 2}^{+}\oplus {\bf 2}^{-}$
\\
$Y_j$
& ${\bf 2}^{+}\oplus{\bf 2}^{+}\oplus {\bf 2}^{-}$
\\
\hline
\end{tabular}
\caption{Irreducible decomposition of the chiral zero-modes and Yukawa couplings.
The upper indices denote the eigenvalue of the diagonal $\mathbb{Z}_2$ and the lower indices denote the eigenvalues of the $D_4$ generators.}
\label{tab:irre_336}
\end{center}
\end{table}

\subsubsection*{Model  without Wilson line}

If the Wilson line is set to zero, 
the Yukawa invariant modular subgroup is enhanced.
The Yukawa term is invariant under $S$
for the vanishing Wilson line model and 
the Yukawa invariant subgroup is enhanced to 
$(\mathbb{Z}_8^{(ST^6)} \times \mathbb{Z}_2^{(S^2)})\rtimes \mathbb{Z}_2^{(T^6)}$.
The character indices of this group and irreducible representations
are summarized in
Table~\ref{tab:character_table_336}.
This group has eight singlets and six doublets.
The three-generation chiral zero-modes are decomposed to three singlets:
\begin{align}
{\bf 3} = {\bf 1}_{+0}\oplus{\bf 1}_{+2}\oplus {\bf 1}_{+1}
\end{align}
where the index represents the eigenvalues of $T^6$ and $S$;
$T^6 {\bf 1}_{\pm j } = \pm {\bf 1}_{\pm j }$ and  $S {\bf 1}_{\pm j } = e^{j \frac{i \pi}{2} } {\bf 1}_{\pm j }$.
The six-generation zero-modes are decomposed into three doublets:
\begin{align}
{\bf 6} = {\bf 2}_{2}\oplus{\bf 2}_{3}\oplus {\bf 2}_{4}.
\end{align}
The representation of the Yukawa is the complex conjugate of that of the six-generation chiral zero-modes:
\begin{align}
\bar {\bf 6} = \bar {\bf 2}_{2}\oplus{\bf 2}_{3}\oplus {\bf 2}_{4} = {\bf 2}_{1}\oplus{\bf 2}_{3}\oplus {\bf 2}_{4}.
\end{align}
Table~\ref{tab:irre_336_without} summarizes the irreducible decomposition of each component.

\begin{table}
\begin{center}
\begin{tabular}{|c | c | } 
\hline
& Representation of $(\mathbb{Z}_8 \times \mathbb{Z}_2) \rtimes \mathbb{Z}_2$
\\
\hline \hline
$\psi^{j,3}$
& ${\bf 1}_{+0}\oplus{\bf 1}_{+2}\oplus {\bf 1}_{+1}$  
\\
$\psi^{j,6}$
& ${\bf 2}_2 \oplus{\bf 2}_3 \oplus {\bf 2}_4$ 
\\
$Y_j$
& ${\bf 2}_1 \oplus{\bf 2}_3 \oplus {\bf 2}_4$
\\
\hline
\end{tabular}
\caption{Irreducible decomposition of the chiral zero-modes and Yukawa couplings without Wilson line.}
\label{tab:irre_336_without}
\end{center}
\end{table}

\begin{table}[t]
\begin{center}
\footnotesize
\begin{tabular}{|c |  c c c c c c c c c c c c c c c |} \hline
&  $h$
& $\chi_{1_{+0}}$ & $\chi_{1_{+1}}$ & $\chi_{1_{+2}}$ & $\chi_{1_{+3}}$ & $\chi_{1_{-0}}$ & $\chi_{1_{-1}}$ & $\chi_{1_{-2}}$ & $\chi_{1_{-3}}$ 
& $\chi_{2_1}$ & $\chi_{2_2}$ & $\chi_{2_3}$ & $\chi_{2_4}$ & $\chi_{2_5}$ & $\chi_{2_6}$ 
\\ 
\hline
\hline
$C_1$ &  1 
& 1 & 1 & 1 & 1 & 1 & 1 & 1 & 1 & 2 & 2 & 2 & 2 & 2 & 2
\\ 
$C_2$ &  2 
& 1 & $-1$ & $1$ & $-1$ & $1$ & $-1$ & $1$ & $-1$ 
& $2$ & $2$ & $-2$ & $-2$ & $2$ & $-2$
\\ 
$C_3$ &  2
& 1 & $-1$ & $1$ & $-1$ & $-1$ & $1$ & $-1$ & $1$ 
& $0$ & $0$ & $0$ & 0 &  $0$ & $0$
\\ 
$C_4$ &  2 
& 1 & $-1$ & $1$ & $-1$ & $1$ &$-1$ & $1$ & $-1$ 
& $-2$ & $-2$ & $2$ & 2  & $2$ & $-2$
\\ 
$C_5$ &  2 
& $1$ & $1$ & $1$ & $1$ &$1$ & $1$ & $1$ & $1$ 
& $-2$ & $-2$ & $-2$ & $-2$ & $2$ & $2$
\\ 
$C_6$ &  2 
& 1 & $1$ & $1$ & $1$ & $-1$ & $-1$ & $-1$ & $-1$ 
& $0$ & $0$ & $0$ & $0$ & $0$ & 0
\\ 
$C_7 $ & 4 
& 1 & $i$ & $-1$ & $-i$ & $1$ & $i$ & $-1$ & $-i$ 
& $0$ & $0$ & $0$ & $0$ & $0$ & 0
\\ 
$C_8$ & 4 
& $1$ & $-1$ & $1$ & $-1$ &$1$ & $-1$ & $1$ & $-1$ 
& $0$ & $0$ & $0$ & $0$ &  $-2$ & $2$  
\\ 
$C_9$ &  4 
& $1$ & $1$ & $1$ & $1$ &$1$ & $1$ & $1$ & $1$ 
& $0$ & $0$ & $0$ & $0$ &  $-2$ & $-2$
\\ 
$C_{10}$  & 4
 & $1$ & $-i$ & $-1$ & $i$ &$1$ & $-i$ & $-1$ & $i$ 
& $0$ & $0$ & $0$ & $0$ & $0$ & $0$
\\ 
$C_{11}$  & 8 
& $1$ & $-i$ & $-1$ & $i$ &$-1$ & $i$ & $1$ & $-i$ 
& $-i\sqrt{2}$ & i$\sqrt{2}$ & $\sqrt{2}$ & $-\sqrt{2}$  &  $0$ &  $0$
\\
$C_{12}$  & 8 
& $1$ & $-i$ & $-1$ & $i$ &$-1$ & $i$ & $1$ & $-i$ 
& $i\sqrt{2}$ & $-i\sqrt{2}$ & $-\sqrt{2}$ & $\sqrt{2}$ &  $0$ &  $0$
\\
$C_{13}$  & 8 
& $1$ & $i$ & $-1$ & $-i$ &$-1$ & $-i$ & $1$ & $i$ 
& $i\sqrt{2}$ & $-i\sqrt{2}$ & $\sqrt{2}$&  $-\sqrt{2}$  & $0$ & $0$
\\ 
$C_{14}$  & 8 
& $1$ & $i$ & $-1$ & $-i$ & $-1$ & $-i$ & $1$ & $i$ 
& $-i\sqrt{2}$ & $i\sqrt{2}$ & $-\sqrt{2}$ & $\sqrt{2}$ & $0$ & $0$
\\
\hline
\end{tabular}
\caption{Character table for the Yukawa invariant modular subgroup which keeps the Yukawa term invariant for the model without Wilson line.}
\label{tab:character_table_336}
\end{center}
\end{table}

\subsubsection*{Comments on the possibility of exceptional elements}

We see if there is an exceptional element that is not covered 
by the generators of $S^2$, $T^6$ and $(ST^6)^2$ ($S$ and $T^6$ for vanishing Wilson line).
Since the modular group of $\{S, T^2\}$ is finite with the order of $768 = 2^{8}\times 3$, 
we can numerically check if each modular transformation 
satisfies the condition \eqref{eq:rhoTrho}. 
In our analysis the group elements of the modular transformation
are obtained with a specific representation e.g., $\rho_{M}$,
so that the group structure should be defined using the largest representation for definiteness. 
In this case, we use the definition for the group element of the modular transformation as
\begin{align}
\label{Eq:rep}
\rho=\rho_{M_1}\oplus\rho_{M_2}\oplus\rho_{M_3}\oplus\rho_Y,
\end{align}
for concrete calculation. 
We confirm that there is no other element which keeps the Yukawa term invariant 
other than the elements covered by $S^2$, $T^6$ and $(ST^6)^2$ ($S$ and $T^6$ for vanishing Wilson line).
The Yukawa invariant modular subgroup is isomorphic to a finite group of 
$\mathbb{Z}_2\times D_4$ ($(\mathbb{Z}_8 \times \mathbb{Z}_2) \rtimes \mathbb{Z}_2$ for vanishing Wilson line).

We note that, although 
$\mathcal{M}$ is 
generated by $S^2$, $T^N$, and $(ST^N)^2$ ($S$ and $T^N$ for vanishing Wilson line), 
the group structure differs depending on the magnetic fluxes in the model, 
since the value of $N$ also differs by models.
In fact, we calculate the group structure for other examples with different magnetic fluxes 
in Appendix \ref{App:examples}, 
and show that various discrete groups appear  
as modular flavor symmetry, e.g., $(\mathbb{Z}_2\times\mathbb{Z}_4)\rtimes \mathbb{Z}_2$ for 
a two-generation model.

\section{Modular Extended Discrete Flavor Symmetry}
\label{Sec:MEDS}

It is known that the magnetized torus model has discrete flavor symmetry.
In this section, we study their relationships and 
consider the full symmetry group.

First we briefly review the conventional discrete flavor symmetry~\cite{Abe:2009vi}.
Suppose that there are chiral zero-modes $\phi^{j_1,M_1}, ... , \phi^{j_\ell, M_\ell}$.
If the greatest common divisor of the generation numbers, 
$g = {\rm g.c.d.}(M_1, ... , M_\ell)$, is greater than 1,
the theory is invariant under the following two operators:
\begin{align}
Z&:\phi^{j,M_k} \rightarrow \omega^j \phi^{j,M_k},
\nonumber
\\
C&:\phi^{j,M_k} \rightarrow \phi^{j + J_k ,M_k},
\end{align}
where $M_k = g J_k$, and $\omega = e^{\frac{2\pi i }{g}}$.
$C$ and $Z$ are represented by $g\times g$ matrices as
\begin{align}
C = 
\begin{pmatrix}
0 & 1 & 0 & \cdots & 0\\
0 & 0 & 1 & \cdots & 0\\
\vdots & \vdots & \vdots & \ddots & \vdots\\
1 & 0 & 0 & \cdots & 0\\ 
\end{pmatrix},
~~
Z=
\begin{pmatrix}
1 & 0  & \cdots & 0\\
0 & \omega &  \cdots & 0\\
\vdots & \vdots & \ddots & \vdots\\
0 & 0 &  \cdots & \omega^{g-1}\\ 
\end{pmatrix},
\end{align}
These two generators satisfy $ZC = \omega CZ$, and there are three $\mathbb{Z}_g$ charges in this model.
Hence this group is isomorphic to $(\mathbb{Z}_g'\times \mathbb{Z}_g^{(Z)} )\rtimes \mathbb{Z}_g^{(C)}$.

We should emphasize that this discrete symmetry is
different from the non-Abelian symmetry originated from the modular subgroup.
The clear difference comes from the fact that the Yukawa couplings are always trivial singlet 
under the conventional flavor symmetry, but not under the modular transformation.

Let $\mathcal{F}$ and $\mathcal{M}$ be the conventional flavor group and the Yukawa invariant modular subgroup, respectively.
As pointed out in \cite{Nilles:2020nnc}, $\mathcal{F}$ and $\mathcal{M}$ are noncommutative with each other.
To see this,
we consider three-generation chiral zero-modes for the purpose of illustration.
The matrix representation of $S^2$ for the three-generation zero-modes is given by \eqref{eq:matrix_3-generation}.
$C$ of $\Delta(27)$ can act on the zero-modes too.
Their three-dimensional representations are given by
\begin{align}
\rho_3(S^2) = 
\begin{pmatrix}
1 & 0 & 0 \\
0 & 0 & 1 \\
0 & 1 & 0 
\end{pmatrix},~~
\rho_3(C) = 
\begin{pmatrix}
0 & 1 & 0 \\
0 & 0 & 1 \\
1 & 0 & 0 \\
\end{pmatrix}.
\end{align}
Therefore $C S^2 \neq S^2 C$.
The sum of the Yukawa invariant modular subgroup and conventional flavor symmetry generates a new group which acts on the effective theory.
A similar idea has been proposed in \cite{Baur:2019kwi, Nilles:2020nnc}.
In the previous works, however, calculation is restricted to a single chiral field,
and a simultaneous transformation of all the components of the model including the Yukawa couplings has not been taken into account.
As pointed out in the previous section, we must use large enough representation 
to identify the group elements of $\mathcal{M}$ correctly. 
The same is true for the modular extension of the flavor symmetry.
To see this, let us consider the model with magnetic fluxes $M_1=M_2 =2$ and $M_3=-4$.
Without the Wilson line, this model has $D_4 \times \mathbb{Z}_2$ conventional flavor symmetry 
and $(\mathbb{Z}_2 \times \mathbb{Z}_4)\rtimes \mathbb{Z}_2$ modular symmetry (see Appendix \ref{App:224}).
$\rho_Y(S)\rho_Y(T^4)\rho_Y(S)^{-1} \neq \rho_Y(C)$ since the Yukawa couplings are the trivial singlets under $\mathcal{F}$,
and $S T^4 S^{-1}$ is not identical to $C$ in this model.
However $\rho_4(S)\rho_4(T^4)\rho_4(S)^{-1} = \rho_4(C)$ for the four-generation zero-mode,
and one may misidentify $ST^4S^{-1} =C$ if one restricts the representation to a single field. 
We need a faithful representation of this combined two groups to avoid such ambiguity.
We provide a complete analysis by use of the largest representation of Eq.\eqref{Eq:rep} for magnetized torus models.

We use $\mathcal{G}$ for denoting this novel group referred to as modular extended flavor group.
Our goal of this section is to analyze the structure of $\mathcal{G}$.
The structure of $\mathcal{G}$ has two possibilities in general.
If $\mathcal{F}$ is not a normal subgroup of $\mathcal{G}$,
this indicates that $\mathcal{F}$ is not the whole flavor symmetry and there is an additional global symmetry hidden in $\mathcal{G}$.
Since in this case we can find ${}^\exists m \in \mathcal{M}$ such that $m \mathcal{F} m^{-1}$ is not identified to $\mathcal{F}$
and the subgroup $m \mathcal{F} m^{-1}$ acts on the Yukawa couplings trivially, this is interpreted as a flavor symmetry, although these two groups are isomorphic.
Otherwise $\mathcal{F}$ denotes the whole flavor symmetry and $\mathcal{M}$ is a subgroup of the automorphism of $\mathcal{F}$ \cite{Nilles:2020nnc}.

Since the representation of the Yukawa couplings is trivial for $\mathcal{F}$, i.e., $\rho_Y(f)={\bf 1}$ for $f \in \mathcal{F}$,
we only need to calculate the algebraic structure for $\rho_M$ 
(the matrix representation for $M$-generation chiral zero-mode) in detail. 
It is convenient to introduce new $M\times M$ matrices $Z'$ and $C'$ as 
\begin{align}
Z' =
\begin{pmatrix}
1 & 0  & \cdots & 0\\
0 & \sigma &  \cdots & 0\\
\vdots & \vdots & \ddots & \vdots\\
0 & 0 &  \cdots & \sigma^{M-1}\\ 
\end{pmatrix},
~~
C' = 
\begin{pmatrix}
0 & 1 & 0 & \cdots & 0\\
0 & 0 & 1 & \cdots & 0\\
\vdots & \vdots & \vdots & \ddots & \vdots\\
1 & 0 & 0 & \cdots & 0\\ 
\end{pmatrix},
\end{align}
where $\sigma= e^{\frac{2\pi i}{M}}$.
These two matrices satisfy the following relations
\begin{align}
\rho_M(S^2)Z' \rho_M(S^{-2}) 
&= 
(Z')^{-1}, \notag \\
\rho_M(S^2)C' \rho_M(S^{-2}) 
&= 
C'^{-1}.
\end{align}
Since $\rho_M(Z)= Z'^{M/g}$ and $\rho_M(C) = C'^{M/g}$,
we obtain
\begin{align}
&S^2 Z S^{-2} =Z^{-1},
\label{eq:conjugate_S^2ZS^2}
\\
&S^2 C S^{-2} = C^{-1}.
\label{eq:conjugate_S^2CS^2}
\end{align}
We find $S^2 \mathcal{F}  S^{-2} \subset \mathcal{F}$.\footnote{
Similar analysis for $S^2$ has also been done in \cite{Nilles:2020nnc}.
Note, however, that the action of $S^2$ on the Yukawa couplings is different from \cite{Nilles:2020nnc}, 
since in our model the Yukawa couplings depend on the Wilson line, which also transforms under 
the modular group (see Eq.~\eqref{eq:mtrwf}).
}
If $M$ is even and $N/M$ is odd, 
Eq.~(\ref{eq:T^N-matrix}) becomes 
\begin{align}
\rho_M(T^N)= 
\begin{pmatrix}
1 & 0 &  \cdots & 0~~\\
0 & -1 &  \cdots & 0~~\\
\vdots & \vdots & \ddots & \vdots~~\\
0 & 0 &  \cdots & -1\\ 
\end{pmatrix}.
\label{eq:T^N_pm}
\end{align}
We obtain
\begin{align}
T^N Z T^{-N} &=Z,
\label{eq:conjugate_TNZTN}
\\
T^N C T^{-N} &= (-1)^{M/g}C = C,
\label{eq:conjugate_TNCTN}
\end{align}
where we note that $M/g$ is always even.\footnote{
We show a precise proof here.
Suppose $M_1,M_2, M_3$ are three integer numbers satisfying $M_3=M_1 +M_2$.
$g$ and $N$ are the greatest common divisor and the least common multiple of these three integers respectively.
We introduce new integer numbers $M_j' = M_j/g$,
then we find $M_1' +M_2' =M_3'$ and $N' =N/g$ is the least common multiple of $M_j'$s.
If ${}^\exists M_i \in \{M_1, M_2, M_3\}$ such that both $N/M_i$ and $M_i/g$ are odd,
$N'/M_i' = N/M_i$ must be odd.
Since $N'$ is even, $M_i'$ must be even. 
This is in contradiction with the assumption.
}
$T^N$ is commutative with the group elements of $\mathcal{F}$.
Using the matrix representation given in \eqref{eq:ST^N^2}, 
we obtain
\begin{align}
\rho_M((ST^N)^2)_{ii'}C'_{i'j'} \rho_M((ST^N)^{-2})_{j'j} &= (-1)^{i-1} \delta^{(M)}_{i, -i' - \frac M 2} \delta_{i',j'-1}
(-1)^{j'-1-\frac M 2} \delta^{(M)}_{j', -j - \frac M 2}
\nonumber
\\
&= (-1)^{i + j-1} \delta_{i,j+1} 
\nonumber
\\
&= (C')^{-1}
\\
\rho_M((ST^N)^2)_{ii'}Z'_{i'j'} \rho_M((ST^N)^{-2})_{j'j} &= (-1)^{i-1}  \delta^{(M)}_{i, -i' - \frac M 2} \sigma^{i'-1} \delta_{i',j'}
(-1)^{j'-1 -\frac M 2}  \delta^{(M)}_{j', -j - \frac M 2}
\nonumber
\\
&=  \sigma^{1-i} \delta_{i,j}
\nonumber
\\
&= (Z' )^{-1}.
\end{align}
Thus we find 
\begin{align}
(ST^N)^2 C(ST^N)^{-2} = C^{-1},
\label{eq:ST^N^2CST^N^-2}
\\
(ST^N)^2 Z (ST^N)^{-2} =  Z^{-1}.
\label{eq:ST^N^2ZST^N^-2}
\end{align}
The above two relations hold even if $M$ is odd or $N/M$ is even, i.e., $\rho_M(T^N)=1$. 
Thus we find that $\mathcal{F}$ is a normal subgroup of $\mathcal{G}$, and $\mathcal{G}$ is written as $\mathcal{F M}$.
Therefore there is no additional flavor symmetry hidden in $\mathcal{G}$.
The intersection of $\mathcal{F}$ and $\mathcal{M}$ is the trivial group, i.e., $\{e\}$, since the Yukawa couplings 
are invariant under $\mathcal{F}$.
We conclude $\mathcal{G}$ is isomorphic to the semidirect product of $\mathcal{F}$ and $\mathcal{M}$:
\begin{align}
\mathcal{G}\simeq \mathcal{F} \rtimes \mathcal{M}.
\label{eq:extention_general}
\end{align}

If the Wilson line is set to zero, 
$\mathcal{M}$ is generated by $\{S,T^N\}$.
Using the matrix representation of $S$ given in \eqref{eq:S-matrix},
we calculate
\begin{align}
\rho_M(S)Z' \rho_M(S^{-1}) 
&= 
C',\\
\rho_M(S)C' \rho_M(S^{-1}) 
&= 
Z'^{-1},
\end{align}
and we obtain
\begin{align}
&S Z S^{-1} = S Z'^{M/g} S^{-1} = C'^{M/g} = C,
\label{eq:conjugate_SZS}
\\
&S C S^{-1} = S C'^{M/g} S^{-1} = (Z'^*)^{M/g} = Z^{-1}.
\label{eq:conjugate_SCS}
\end{align}
Therefore we find $S \mathcal{F}  S^{-1} \subset \mathcal{F}$.
In addition, 
there is a parity symmetry $P$ which acts on the wave functions as
\begin{align}
P: \phi^{j,M_k} \rightarrow \phi^{M_k -j, M_k}.
\label{eq:PinF}
\end{align}
and trivially acts on the Yukawa couplings, i.e., $P\in \mathcal{F}$.
$\mathcal{F}$ is generated by $C, Z$ and $P$.
Eq.~\eqref{eq:PinF} is nothing but the action of $S^2$ given in Eq.~\eqref{eq:S^2-matrix}.
Actually the parity operator $P$ is understood as an element of $\mathcal{M}$; $P \in \mathcal{M}$.
Since $Y_{ijk} = Y_{M_1-i, M_2-j, |M_3|-k}$ for vanishing Wilson line,
the action of $S^2$ on the Yukawa couplings is given as 
\begin{align}
S^2~:~Y_{ijk} \rightarrow Y_{M_1-i, M_2-j, |M_3|-k}=Y_{ijk}.
\end{align}
Therefore $P$ is identical to $S^2$ for the vanishing Wilson line%
\footnote{This result is the same as the result of \cite{Nilles:2020nnc}.}
($S^2$ as a generalization of $P$ for nonvanishing Wilson line).
$S^2$ is the unique element except for the identity that keeps the Yukawa couplings invariant in $\mathcal{M}$.
$S^2 = P$ is a center of $\mathcal{M}$, which means $\mathcal{M} P \mathcal{M}^{-1} = P$.
Thus $\mathcal{F}$ is still a normal subgroup of $\mathcal{G}$, and $\mathcal{M}$ is an automorphism of $\mathcal{F}$.
We introduce $\mathcal{F}'$ as a subgroup of $\mathcal{F}$ generated by $C$ and $Z$,
and $\mathcal{G}$ is written as their semidirect product:
\begin{align}
\mathcal{G} \simeq \mathcal{F}' \rtimes \mathcal{M}.
\end{align}

We consider a concrete example in the following subsection for illustration purposes.

\subsection{Modular extended flavor symmetry in three-generation model} 

Here we consider the model of $M_1 =M_2 = 3$ and $M_3=-6$.

\subsubsection*{Model with Wilson line}

First, we consider model with nonvanishing Wilson line.
In this case we have $D_4 \times \mathbb Z_2$ modular symmetry and $\Delta(27)$ for flavor symmetry.
We use fifteen-dimensional representation $\rho_{3} \oplus \rho_{-6} \oplus \rho_Y$ to construct the whole group since there are three- and six-generation chiral zero-modes and 6 Yukawa couplings.
The generators of the modular symmetry is given by
\begin{align}
&\rho_{15}(S^2) = 
\begin{pmatrix}
1 & 0 &0\\
0 & 0 &1\\
0 & 1 &0\\
\end{pmatrix}
\oplus
\begin{pmatrix}
1 & 0 &0 & 0 & 0 & 0\\
0 & 0 &0 & 0 & 0 & 1\\
0 & 0 &0 & 0 & 1 & 0\\
0 & 0 &0 & 1 & 0 & 0\\
0 & 0 &1 & 0 & 0 & 0\\
0 & 1 &0 & 0 & 0 & 0\\
\end{pmatrix}
\oplus
\begin{pmatrix}
1 & 0 &0 & 0 & 0 & 0\\
0 & 0 &0 & 0 & 0 & 1\\
0 & 0 &0 & 0 & 1 & 0\\
0 & 0 &0 & 1 & 0 & 0\\
0 & 0 &1 & 0 & 0 & 0\\
0 & 1 &0 & 0 & 0 & 0\\
\end{pmatrix}
\end{align}
\begin{align}
&\rho_{15}(T^6) = 
1_{3\times 3}
\oplus
\begin{pmatrix}
1 & 0 &0 & 0 & 0 & 0\\
0 & -1 &0 & 0 & 0 & 0\\
0 & 0 &1& 0 & 0 & 0\\
0 & 0 &0 & -1 & 0 & 0\\
0 & 0 &0 & 0 & 1 & 0\\
0 & 0 &0 & 0 & 0 & -1\\
\end{pmatrix}
\oplus
\begin{pmatrix}
1 & 0 &0 & 0 & 0 & 0\\
0 & -1 &0 & 0 & 0 & 0\\
0 & 0 &1& 0 & 0 & 0\\
0 & 0 &0 & -1 & 0 & 0\\
0 & 0 &0 & 0 & 1 & 0\\
0 & 0 &0 & 0 & 0 & -1\\
\end{pmatrix}
\end{align}
\begin{align}
&\rho_{15}((ST^6)^2) = 
\begin{pmatrix}
1 & 0 & 0\\
0 & 0 & 1\\
0 & 1 & 0\\
\end{pmatrix}
\oplus
\begin{pmatrix}
0 & 0 &0 & -1 & 0 & 0\\
0 & 0 &1 & 0 & 0 & 0\\
0 & -1 &0& 0 & 0 & 0\\
1 & 0 &0 & 0 & 0 & 0\\
0 & 0 &0 & 0 & 0 & -1\\
0 & 0 &0 & 0 & 1 & 0\\
\end{pmatrix}
\oplus
\begin{pmatrix}
0 & 0 &0 & -1 & 0 & 0\\
0 & 0 &1 & 0 & 0 & 0\\
0 & -1 &0& 0 & 0 & 0\\
1 & 0 &0 & 0 & 0 & 0\\
0 & 0 &0 & 0 & 0 & -1\\
0 & 0 &0 & 0 & 1 & 0\\
\end{pmatrix}.
\end{align}
where the first $3\times 3$ matrices denote representation for three-generation chiral zero-modes, 
and the second one is for six-generation chiral zero-modes.
The last one acts on the Yukawa couplings.
The conventional flavor group is generated by
\begin{align}
&\rho_{15}(C) = 
\begin{pmatrix}
0 & 1 &0\\
0 & 0 &1\\
1 & 0 &0\\
\end{pmatrix}
\oplus
\begin{pmatrix}
0 & 0 &1 & 0 & 0 & 0\\
0 & 0 &0 & 1 & 0 & 0\\
0 & 0 &0 & 0 & 1 & 0\\
0 & 0 &0 & 0 & 0 & 1\\
1 & 0 &0 & 0 & 0 & 0\\
0 & 1 &0 & 0 & 0 & 0\\
\end{pmatrix}
\oplus
1_{6\times 6}
\\
&\rho_{15}(Z) = 
\begin{pmatrix}
1 & 0 &0\\
0 & \omega &0\\
0 & 0 & \omega^2\\
\end{pmatrix}
\oplus
\begin{pmatrix}
1 & 0 &0 & 0 & 0 & 0\\
0 & \omega &0 & 0 & 0 & 0\\
0 & 0 &\omega^2& 0 & 0 & 0\\
0 & 0 &0 & 1 & 0 & 0\\
0 & 0 &0 & 0 & \omega & 0\\
0 & 0 &0 & 0 & 0 & \omega^2\\
\end{pmatrix}^*
\oplus
1_{6\times 6}.
\end{align}
$\rho_{15}(Z)$ has the conjugate representation for the six-generation chiral zero-mode since $M_3$ is negative.
The irreducible decomposition of this group is summarized in Table~\ref{tab:MEFS_336}.
\begin{table}[t]
\begin{center}
\begin{tabular}{|c | c |} 
\hline
& $\Delta(27)$
\\
\hline \hline
$\phi^{j,3}$
& ${\bf 3}$
\\
$\phi^{j,6}$
& $ 2\times \bar {\bf 3}$
\\
$Y_j$
& $6\times {\bf 1}$
\\
\hline
\end{tabular}
\caption{Irreducible decomposition of the chiral zero-modes and Yukawa couplings
under the conventional flavor symmetry $\Delta(27)$ \cite{Abe:2009vi}.
}
\label{tab:MEFS_336}
\end{center}
\end{table}

The following relations can be shown:
\begin{align}
T^6 C T^6 &=C,
\nonumber
\\
T^6 Z T^6 &=Z,
\nonumber
\\
S^2 C S^2 &= C^2,
\nonumber
\\
S^2 Z S^2 &=Z^2,
\nonumber
\\
(T^6S)^2 C (T^6S)^{-2} &= C^2,
\nonumber
\\
(T^6S)^2 Z (T^6S)^{-2} &= Z^2.
\end{align}
These are equivalent to \eqref{eq:conjugate_S^2ZS^2}, \eqref{eq:conjugate_S^2CS^2}, \eqref{eq:conjugate_TNZTN},
\eqref{eq:conjugate_TNCTN},  \eqref{eq:ST^N^2CST^N^-2}, and \eqref{eq:ST^N^2ZST^N^-2}.
Thus the conventional flavor group $\mathcal{F}$ is the normal subgroup of the novel group $\mathcal{G}$.
The intersection of $\mathcal{F}$ and $\mathcal{M}$ consists only of the identity since the action of $\mathcal{F}$ on the Yukawa couplings is always trivial.
We conclude $\mathcal{G}$ is the semidirect product of $\mathcal{F}$ and $\mathcal{M}$:
\begin{align}
\mathcal{G}\simeq \mathcal{F}\rtimes \mathcal{M} = \Delta(27) \rtimes (D_4\times \mathbb{Z}_2).
\end{align}
This is the modular extension of the flavor group for this three-generation model.

\subsubsection*{Model without Wilson line}

Without the Wilson line, we have additional generators $S$.
The matrix representation of $S$ is given by
\begin{align}
&\rho_{15}(S) = \frac 1 {\sqrt{3}}
\begin{pmatrix}
1 & 1 &1\\
1 & \omega & \omega^2\\
1 & \omega^2 & \omega\\
\end{pmatrix}
\oplus
\frac 1 {\sqrt{6}}
\begin{pmatrix}
1 & 1 &1 & 1 & 1 & 1\\
1 & \eta &\eta^2 & -1 & \eta^4 &\eta^5\\
1 & \eta^2 &\eta^4 & 1 & \eta^2 &\eta^4\\
1 & -1 &1 & -1 & 1 &-1\\
1 & \eta^4 &\eta^2 & 1 & \eta^4 &\eta^2\\
1 & \eta^5 &\eta^4 & -1 & \eta^2 &\eta^1\\
\end{pmatrix}^*
\oplus
\frac 1 {\sqrt{6}}
\begin{pmatrix}
1 & 1 &1 & 1 & 1 & 1\\
1 & \eta &\eta^2 & -1 & \eta^4 &\eta^5\\
1 & \eta^2 &\eta^4 & 1 & \eta^2 &\eta^4\\
1 & -1 &1 & -1 & 1 &-1\\
1 & \eta^4 &\eta^2 & 1 & \eta^4 &\eta^2\\
1 & \eta^5 &\eta^4 & -1 & \eta^2 &\eta^1\\
\end{pmatrix},
\nonumber
\end{align}
In addition we have $P \in \mathcal{F}$, and $\mathcal{F} \simeq \Delta(54)$. 
We note that $P$ is identical to $S^2$ since $Y_{jk\ell} = Y_{-i\, -j\, -\ell}$ as we denoted in the previous section.
The conjugation by $S$ is given by
\begin{align}
S ZS^{-1} &= C,
\\
S C S^{-1} &= Z^2.
\end{align}
These are equivalent to \eqref{eq:conjugate_SZS} and \eqref{eq:conjugate_SCS}.
$\mathcal{F}' \simeq \Delta(27)$ and $\mathcal{M} \simeq (\mathbb{Z}_8 \times \mathbb{Z}_2) \rtimes \mathbb{Z}_2$.
Therefore $\mathcal{G}$ is written as
\begin{align}
\mathcal{G} \simeq \mathcal{F}' \rtimes \mathcal{M}
\simeq \Delta(27) \rtimes ((\mathbb{Z}_8 \times \mathbb{Z}_2) \rtimes \mathbb{Z}_2).
\end{align}
Irreducible decomposition of the three-generation chiral zero-mode is given 
by a three-dimensional representation since it is ${\bf 3}$ in $\Delta(27)$.
The six-generation chiral zero-modes are six-dimensional representation of $\mathcal{G}$.
The Yukawa couplings are decomposed to three two-dimensional representations,
since they are trivial representation in $\Delta(27)$.

\section{Conclusion}

We have investigated the modular symmetry of the magnetized torus.
The modular group is isomorphic to $SL(2,\mathbb Z)/\mathbb{Z}_2$ and it is an infinite group.
For the heterotic orbifold, 
the modular group can act on its effective action and it is invariant under the whole group.
However, for magnetized torus, the situation is different.
When the magnetic fluxes turn on
effective action is no longer invariant under the whole modular group,
but is invariant under its specific subgroup $\mathcal{M}$, which we refer to modular flavor symmetry.
We have shown this group consists of $S^2$, $T^N$, and $(ST^N)^2$, where $N$ is the least common multiple of the generation numbers in general.
These elements are noncommutative and generate non-Abelian groups. 
This group is enhanced for the case of vanishing Wilson line, and the theory (the Yukawa term) becomes $S$ invariant.
We show several examples of constructions of this Yukawa invariant subgroups.
These subgroups are isomorphic to finite groups, such as 
$D_4\times \mathbb{Z}_2$ and $(\mathbb{Z}_8 \times \mathbb{Z}_2)\rtimes \mathbb{Z}_2$.
We find the group structures depend on the chiral spectrum and we can realize various finite groups as subgroups of the modular group.
The modular flavor symmetry consists of several $\mathbb{Z}_{2}, \mathbb{Z}_{4}$ and $\mathbb{Z}_{8}$.
Such discrete groups are utilized for solving the flavor puzzles \cite{Ishimori:2008gp}.

It is known that the magnetized torus model has conventional flavor symmetry $\mathcal{F}$.
This flavor symmetry includes the parity symmetry in terms of the extra dimension if the Wilson line vanishes. 
Although the modular group and the conventional flavor group are different, 
we have found that the parity operator can be interpreted as $S^2$ in the modular symmetry.
We have investigated modular extension of conventional flavor symmetry in detail.
They are noncommutative with each other and enlarge the group of the symmetry.
Such an extension of the flavor symmetry has been studied in \cite{Nilles:2020nnc}.
However, we have extended the analysis to modular transformation of the Yukawa terms, which is important to correctly analyze the symmetry of the theory.
We have found there is no additional flavor symmetry hidden in the novel group $\mathcal{G}$ (modular extended flavor group).
Therefore, as pointed out in \cite{Nilles:2020nnc}, 
the conventional flavor group $\mathcal{F}$ is a normal subgroup of $\mathcal{G}$ 
and $\mathcal{M}$ is a subgroup of the automorphism of $\mathcal{F}$. 
In addition, we have found that $\mathcal{G}$ is isomorphic to the semidirect product of modular and the conventional flavor group for nonvanishing Wilson line,
because the Yukawa couplings form a faithful representation of $\mathcal{G}$.
For the vanishing Wilson line, there is a nontrivial common element between $\mathcal{F}$ and $\mathcal{M}$, which is $S^2$ in $\mathcal{M}$.
This is identical to $P$ in $\mathcal{F}$.
Thus $\mathcal{G}$ is not the semidirect product of $\mathcal{F}$ and $\mathcal{M}$, but the semidirect product of its subgroup $\mathcal{F}'$, which is generated by $Z$ and $C$.

Our study is based on a field theory analysis of the magnetized torus model
which is the low-energy effective theory of type II string theory.
Taking into account more stringy effects, e.g., vertex operator, local supersymmetry, or the Green-Schwartz-like anomaly cancellation mechanism, 
modular properties of fields and couplings may change.
Pursuing this possibility is certainly interesting, but it is beyond the scope in the present paper.

\section*{Acknowledgments}

We would like to thank Patrick K.S. Vaudrevange for helpful comments about
general roles of the automorphism of finite groups and its phenomenological application.
H.~O. is supported in part by JSPS KAKENHI Grants 
No.\,17K14309 and No.\,18H03710.

\appendix

\section{More examples of Yukawa invariant modular subgroups}
\label{App:examples}

We calculate more examples of Yukawa invariant modular subgroups in this appendix.
We study models similar to the model studied in Section 3;
the models contains three gauge groups $SU(N_1) \times SU(N_2) \times SU(N_3)$ and three types of bifundamental chiral zero-modes.
Their generation numbers are given by $M_1$, $M_2$ and $M_3$.
They satisfy $M_1 + M_2 +M_3 =0.$

\subsection{224 model}
\label{App:224}

Let $M_1=M_2=2$ and $M_3=-4$.
In this case, there are two two-generation chiral zero-modes and one four-generation chiral zero-mode.
The matrix representations of the generators of the modular group for the two-generation chiral zero-modes are given by
\begin{align}
\rho_2(S) =\frac{1}{\sqrt{2}} 
\begin{pmatrix}
1 & 1\\
1 & -1
\end{pmatrix},~~
\rho_2(T) = 
\begin{pmatrix}
1 & 0\\
0 & i
\end{pmatrix},
\label{eq:mat_S_2}
\end{align}
and for $M = -4$, the matrix representations of $S$ and $T$ are given by
\begin{align}
\rho_{-4}(S)= \frac 1 {\sqrt{4}}
\begin{pmatrix}
1 & 1 &1 & 1\\
1 & i & -1 & -i\\
1 & -1 & 1 & -1\\
1 & -i & -1 & i
\end{pmatrix}^*,
~~
\rho_{-4}(T)=
\begin{pmatrix}
1 & 0 & 0 & 0\\
0 & e^{\frac{\pi i }4} &0 &0\\
0 & 0 & -1 & 0\\
0 & 0 & 0 & e^{\frac{\pi i }4}
\end{pmatrix}^*.
\label{eq:M=4_matrix}
\end{align}
where the complex conjugate is required since $M_3$ is negative.

\subsubsection*{Model with Wilson line}

First we investigate the model with nonvanishing Wilson line.
In this case the Yukawa couplings are classified to four values:
\begin{align}
Y_{0}(\tau) = Y_{000}= Y_{112},~~
Y_1(\tau) = Y_{101}= Y_{013},~~
Y_2(\tau)= Y_{110}= Y_{002},~~
Y_3(\tau)= Y_{011} = Y_{103},
\end{align}
and these four $Y_j$ form a four-dimensional representation of the modular group.
The Yukawa invariant subgroup is generated by $S^2, T^4$ and $(S T^4)^2$.
They satisfy the following equations:
\begin{align}
(S^2)^2 = (T^4)^2= ((S T^4)^2)^2 =1.
\end{align}
Hence they correspond to $\mathbb{Z}_2$.
They are commutative with each other and the group is isomorphic to $\mathbb{Z}_2 \times \mathbb{Z}_2 \times \mathbb{Z}_2$.
We also check that there is no extra element which keeps the Yukawa term invariant but can not be 
generated by $S^2$, $T^4$, and $(ST^4)^2$
in the group generated by $\rho_M(S)$ and $\rho_M(T)$, which consists of $3072 = 2^{10}\times 3$ elements.
Irreducible decomposition of the representations is summarized in Table~\ref{tab:irreducible_224}.

\begin{table}[th]
\begin{center}
\begin{tabular}{|c|c|} \hline
 & $\mathbb{Z}_2 \times \mathbb{Z}_2 \times \mathbb{Z}_2 $
\\
\hline \hline
$\phi^{j,2}$ 
& $2\times {\bf 1}_{++}$
\\ 
$\phi^{j,4}$ 
& ${\bf 1}_{---} \oplus {\bf 1}_{+--} \oplus {\bf 1}_{++-} \oplus {\bf 1}_{+++}  $ 
\\
$Y_j$ 
& ${\bf 1}_{---} \oplus {\bf 1}_{+--} \oplus {\bf 1}_{++-} \oplus {\bf 1}_{+++}  $ 
\\
\hline
\end{tabular}
\caption{Irreducible decomposition of the fields and Yukawa couplings for model with Wilson line. 
The indices of ${\bf 1}_{jk\ell}$ denote the eigenvalues of $\mathbb{Z}_2^{(S^2)}, \mathbb{Z}_2^{(T^4)}$ and
$\mathbb{Z}_2^{(ST^4)^2}$ respectively.
}
\label{tab:irreducible_224}
\end{center}
\end{table}

\subsubsection*{Model without Wilson line}

For the vanishing Wilson line model,
the Yukawa invariant modular group is enhanced.
This group has 16 elements and it contains two $\mathbb{Z}_2$ and one $\mathbb{Z}_4$.
The $\mathbb{Z}_4$ corresponds to $S$, and the two $\mathbb{Z}_2$ correspond to $T^4$ and $(S T^4)^2$.
Therefore, this group is generated by $S$ and $T^4$.
They satisfy the relations
\begin{align}
T^4 S T^{-4} &= S^3 (ST^4)^2,
\nonumber
\\
S (ST^4)^2  S^{-1} &= (ST^4)^2,
\nonumber
\\
T^4 (ST^4)^2 T^{-4} &= (ST^4)^2,
\end{align}
and these mean that the subgroup generated by $S$ and
$(ST^4)^2$ is a normal subgroup of the whole group.
The group generated by these matrices is isomorphic to
$(\mathbb{Z}_2^{(ST^4)^2} \times \mathbb{Z}_4^{(S)})  \rtimes \mathbb{Z}_2^{(T^4)}$.
This is the modular symmetry of the Yukawa term without the Wilson line.
Irreducible decomposition of the representations is summarized in Table~\ref{tab:irreducible_224_w/oWL}.

\begin{table}[t]
\begin{center}
\begin{tabular}{|c|c|} \hline
 & $(\mathbb{Z}_2 \times \mathbb{Z}_4) \rtimes \mathbb{Z}_2 $
\\
\hline \hline
$\phi^{j,2}$ 
& ${\bf 1}_{+0} \oplus {\bf 1}_{+2} $ 
\\ 
$\phi^{j,4}$ 
& ${\bf 1}_{+0} \oplus {\bf 1}_{-3} \oplus{\bf 2^*}$ 
\\
$Y_j$ 
& ${\bf 1}_{+0} \oplus {\bf 1}_{-1} \oplus{\bf 2}$ 
\\
\hline
\end{tabular}
\caption{Irreducible decomposition of the fields and Yukawa couplings for the model without the Wilson line. 
The indices of ${\bf 1}_{jk}$ in the right column denote the eigenvalues of $ \mathbb{Z}_2^{(T^4)}$ and $\mathbb{Z}_4^{(S)} $ respectively.
}
\label{tab:irreducible_224_w/oWL}
\end{center}
\end{table}

\subsubsection*{Modular extended discrete flavor symmetry}

This model has $D_4$ flavor symmetry in general
and $D_4 \times \mathbb{Z}_2$ flavor symmetry for the vanishing Wilson line model.
These $D_4$ and the Yukawa invariant modular subgroups are noncommutative.
As shown in Section \ref{Sec:MEDS}, we can obtain modular extended flavor symmetry.
The Yukawa invariant modular subgroup generators are given by
\begin{align}
&\rho_{10}(S^2) =
1_{2\times 2}
\oplus
\begin{pmatrix}
1 & 0 &0 & 0\\
0 & 0 & 0 & 1\\
0 & 0 & 1 & 0\\
0 & 1 & 0 & 0
\end{pmatrix}
\oplus
\begin{pmatrix}
1 & 0 &0 & 0\\
0 & 0 & 0 & 1\\
0 & 0 & 1 & 0\\
0 & 1 & 0 & 0
\end{pmatrix}
,
\nonumber
\\
&\rho_{10}(T^4) =
1_{2\times 2}\oplus
\begin{pmatrix}
 1 & 0 & 0 & 0 \\
 0 & -1 & 0 & 0 \\
 0 & 0 & 1 & 0 \\
 0 & 0 & 0 & -1 
\end{pmatrix}
\oplus
\begin{pmatrix}
 1 & 0 & 0 & 0 \\
 0 & -1 & 0 & 0 \\
 0 & 0 & 1 & 0 \\
 0 & 0 & 0 & -1 
\end{pmatrix},
\\
&\rho_{10}((S T^4)^2)=
1_{2\times 2}
\oplus
\begin{pmatrix}
0 & 0 & 1 & 0\\
0 & -1 & 0 &0\\
1 & 0 & 0 & 0\\
0 & 0 & 0 & -1
\end{pmatrix}
\oplus
\begin{pmatrix}
0 & 0 & 1 & 0\\
0 & -1 & 0 &0\\
1 & 0 & 0 & 0\\
0 & 0 & 0 & -1
\end{pmatrix}.
\nonumber
\end{align}
The flavor group generators are similarly given by
\begin{align}
&\rho_{10}(C) =
\begin{pmatrix}
0 & 1  \\
1 & 0  
\end{pmatrix}
\oplus
\begin{pmatrix}
 0 & 0 & 1 & 0 \\
 0 & 0 & 0 & 1 \\
 1 & 0 & 0 & 0 \\
 0 & 1 & 0 & 0 
\end{pmatrix}
\oplus
1_{4\times 4}
,
\nonumber
\\
&\rho_{10}(Z) =
\begin{pmatrix}
1 & 0  \\
0 & -1 
\end{pmatrix}
\oplus
\begin{pmatrix}
 1 & 0 & 0 & 0 \\
 0 & -1 & 0 & 0 \\
 0 & 0 & 1 & 0 \\
 0 & 0 & 0 & -1 
\end{pmatrix}
\oplus
1_{4\times 4}.
\end{align}
Their irreducible decomposition is summarized in Table~\ref{tab:irr_FM_224}.
\begin{table}[t]
\begin{center}
\footnotesize
\begin{tabular}{|c|c|} \hline
& Representation of $D_4$ 
\\ 
\hline \hline
$\phi^{j,2}$ 
& ${\bf 2}$  
\\
$\phi^{j,4}$ 
& ${\bf 1}_{++} \oplus {\bf 1}_{+-} \oplus {\bf 1}_{-+} \oplus {\bf 1}_{--} $  
\\
$Y_j$ 
& $4\times {\bf 1}_{++}$ 
\\ \hline  
\end{tabular}
\caption{Irreducible representation of the conventional flavor symmetry,}
\label{tab:irr_FM_224}
\end{center}
\end{table}

It is easy to show that $\mathcal{M}$ is commutative with all the generators of $\mathcal{F}$.
This is because $C^{-1}$ and $Z^{-1}$ are the same as $C$ and $Z$.
Thus, the whole group $\mathcal{G}$ is isomorphic to the direct product of $\mathcal{F}$ and $\mathcal{M}$.
We find
\begin{align}
G \simeq D_4 \times (\mathbb{Z}_2)^3.
\end{align}

Without the Wilson line, the modular group is enhanced to $(\mathbb{Z}_2\times \mathbb{Z}_4)\rtimes \mathbb{Z}_2$.
It is generated by $S$ and $T^4$.
The 10-dimensional representation of $S$ is written as
\begin{align}
&\rho_{10}(S) =
\frac 1 {\sqrt{2}}
\begin{pmatrix}
1 & 1  \\
1 & -1  
\end{pmatrix}
\oplus
\frac 1 {\sqrt{4}}
\begin{pmatrix}
1 & 1 &1 & 1\\
1 & i & -1 & -i\\
1 & -1 & 1 & -1\\
1 & -i & -1 & i
\end{pmatrix}^*
\oplus
\frac 1 {\sqrt{4}}
\begin{pmatrix}
1 & 1 &1 & 1\\
1 & i & -1 & i^3\\
1 & -1 & 1 & -1\\
1 & -i & -1 & i
\end{pmatrix}
\end{align}
We also have an extra $\mathbb{Z}_2$ symmetry,
which acts on the chiral zero-modes as $\psi^{j,M} \rightarrow \psi^{-j,M}$.
This $\mathbb{Z}_2$ is denoted by $P$ and its matrix representation is the same as that of $S^2$.
The following relations hold:
\begin{align}
SC S^{-1} &= Z
\\
S Z S^{-1} &= C
\\
T^4 C(T^4)^{-1} &= C
\\
T^4 Z(T^4)^{-1} &= Z
\end{align}
These are nothing but \eqref{eq:conjugate_TNZTN}, \eqref{eq:conjugate_TNCTN}, \eqref{eq:conjugate_SZS} and \eqref{eq:conjugate_SCS}.
These relations mean the flavor symmetry group is a normal subgroup of the whole symmetry group.
The intersection of the $D_4$ and the modular group is a trivial subgroup: $D_4 \cap \mathcal{M}= \{ e\}$.
Therefore the whole symmetry group is semidirect product of $D_4$ and $\mathcal{M}$:
\begin{align}
G\simeq D_4 \rtimes  ((\mathbb{Z}_2\times\mathbb{Z}_4)\rtimes \mathbb{Z}_2).
\end{align}
This is the full symmetry of the effective action.
Since this group is denoted by the (semi)direct product of the groups,
its order is $128 = 8\times 16$.

\subsection{246 model}

Here we consider the model with $M_1 =2, M_2 =4$ and $M_3 =-6$.
The matrix representation of the modular transformation is already given in the former subsections. 
Since $g.c.d.(M_1, M_2,|M_3|) = 2$, we have $D_4$ discrete flavor symmetry for nonzero Wilson line models and $D_4\times \mathbb{Z}_2$ for the vanishing Wilson line.
Yukawa couplings are classified into 12 values:
\begin{align}
&Y_0 = Y_{000} = Y_{123},~~
&Y_1 = Y_{035} = Y_{112},~~~~
&Y_2 = Y_{024} = Y_{101},~~
&Y_3 = Y_{013} = Y_{130}
\nonumber
\\
&Y_4 = Y_{002} = Y_{125},~~
&Y_5 = Y_{031} = Y_{114},~~~~
&Y_6 = Y_{021} = Y_{103},~~
&Y_7 = Y_{015} = Y_{132}
\nonumber
\\
&Y_8 = Y_{004} = Y_{121},~~
&Y_9 = Y_{033} = Y_{110},~~~~
&Y_{10} = Y_{022} = Y_{105},~~
&Y_{11} = Y_{011} = Y_{134}.
\end{align}
The other three-point couplings are prohibited by $\mathbb{Z}_2$ charge.
We obtain 12-dimensional representation of the modular group.
This Yukawa term is not invariant under the whole modular group.
We construct its subgroup under which the Yukawa term is invariant.
If the Wilson line is zero, this subgroup consists of 16 elements.
This group is isomorphic to $(\mathbb{Z}_2 \times \mathbb{Z}_4)\rtimes \mathbb{Z}_2$.
All elements are commutative with each other.
If the Wilson line is not zero, it is not invariant under $S$, but $S^2$, and the group is broken to $\mathbb{Z}_2\times\mathbb{Z}_2 \times \mathbb{Z}_2$.

\subsection{123 model}

Here we consider the model of $M_1 = 1, M_2 =2$, and $M_3 =-3$.
In this model, there are one one-generation chiral superfield, one two-generation chiral superfield, and three-generation chiral superfield.
Their matrix representations of the modular transformation have been given already.
In addition, we have six Yukawa couplings for general Wilson line case.
Their modular transformation is the same as that of six-dimensional chiral zero-mode.
If the Wilson line is zero, we have $\mathbb{Z}_2$ parity flavor symmetry.
We use 11-dimensional representation to construct the Yukawa invariant modular subgroup:
$\rho_{\bf 11} = \rho_2\oplus \rho_3^* \oplus \rho_Y = \rho_2\oplus \rho_3^* \oplus \rho_6.$
We find that they generate a finite group whose order is 768.

The Yukawa invariant modular subgroup is generated by $S$ and $T^6$.
The subgroup consists of 32 elements.
This group is the same as that of the $336$ model.
This group is isomorphic to $(\mathbb{Z}_8\times \mathbb{Z}_2)\rtimes \mathbb{Z}_2$.
If nonzero Wilson line is turned on, $S$ is no longer an element of the Yukawa invariant modular subgroup.
The modular subgroup is broken to $D_4\times \mathbb{Z}_2$.

\end{document}